\documentclass[a4paper,11pt]{article}
\pdfoutput=1
\usepackage{graphicx,epsfig}
\usepackage{grffile}
\usepackage{amsmath,amssymb}
\usepackage{array}
\usepackage[usenames, dvipsnames]{color}
\usepackage{slashed}
\usepackage[normalem]{ulem}
\usepackage{jheppub}
\usepackage{url}

\numberwithin{equation}{section}

\newcommand{\ord}[1]{\mathcal{O}\left({#1}\right)}

\title{\Large\bf\boldmath
Visible neutrino decay in the light of appearance and disappearance long-baseline experiments
}
\date{\today}
\author[a]{Alberto~M. Gago}
\author[b]{Ricardo~A. Gomes}
\author[b]{Abner~L.~G. Gomes}
\author[a]{Joel Jones-P\'erez}
\author[c]{Orlando~L.~G. Peres}

\affiliation[a]{Secci\'on F\'isica, Departamento de Ciencias, Pontificia Universidad Cat\'olica del Per\'u, \\ Apartado 1761, Lima, Peru}
\affiliation[b]{Instituto de F\'isica, Universidade Federal de Goi\'as, 74001-970, Goi\^ania, GO, Brazil}
\affiliation[c]{Instituto de F\'isica Gleb Wataghin, UNICAMP, 13083-859, Campinas, SP, Brazil}

\emailAdd{agago@pucp.edu.pe}
\emailAdd{ragomes@ufg.br}
\emailAdd{abnerleonel.gadelha@gmail.com}
\emailAdd{jones.j@pucp.edu.pe}
\emailAdd{orlando@ifi.unicamp.br}

\abstract{
We investigate the present constraints from MINOS and T2K experiments for the neutrino decay scenario induced by non-diagonal
couplings of Majorons to neutrinos. As novelty, on top of the typical invisible decay prescription, we add the contribution of visible decay, where final products can be observed. This new effect depends on the nature of the neutrino - Majoron coupling, which can be of scalar or pseudoscalar type.
Using the combination of disappearance data from MINOS and disappearance and appearance data from T2K, for normal ordering, we constrain the decay parameter $\alpha\equiv E\,\Gamma$ for the heaviest neutrino, where $E$ and $\Gamma$ are the neutrino energy and width, respectively. We find that when considering visible decay within appearance data, one can improve current neutrino long-baseline constraints up to $\alpha< \ord{10^{-5}}$~eV$^2$, at 90\% C.L., for both kinds of couplings, which is better by one order of magnitude compared to previous bounds.
}

\begin{document}

\maketitle
\flushbottom

\section{Introduction}

Experiments carried out in the last two decades have categorically confirmed the neutrino oscillation phenomenon as the mechanism responsible for the flavour transitions observed in the neutrino fluxes coming from the sun~\cite{Cleveland:1998nv,Abdurashitov:1999zd,Altmann:2000ft,Fukuda:2001nj,Ahmad:2002ka,Davis:2003xx}, the atmosphere~\cite{Fukuda:1998mi,Ambrosio:2001je,Kajita:2016vhj}, reactors~\cite{Araki:2004mb,An:2012eh,Abe:2011fz,Ahn:2012nd} and accelerators~\cite{Adamson:2007gu,Abe:2012gx,Adamson:2013whj,Adamson:2017qqn}. These observations confirm that at least two neutrinos are massive, such that two independent mass differences $\Delta m^2_{ij}=m^2_i-m^2_j\;(i,j=1,\,2,\,3)$ are non-zero~\cite{GonzalezGarcia:2002dz,McDonald:2004rx,Kajita:2016vhj}. In addition, interaction and mass eigenstates are related through the PMNS lepton mixing matrix~\cite{Pontecorvo:1957cp,Maki:1962mu}, defined in terms of three mixing angles and one CP-violation phase $\delta_{\rm CP}$:
\begin{equation}
\label{eq:PMNS}
U_{\rm PMNS}=\left(\begin{array}{ccc}
c_{12}c_{13} & s_{12}c_{13} & s_{13}e^{-i\delta_{\rm CP}} \\
-s_{12}c_{23}-c_{12}s_{23}s_{13}e^{i\delta_{\rm CP}}
& c_{12}c_{23}-s_{12}s_{23}s_{13}e^{i\delta_{\rm CP}} & s_{23}c_{13} \\
s_{12}s_{23}-c_{12}c_{23}s_{13}e^{i\delta_{\rm CP}}
& -c_{12}s_{23}-s_{12}c_{23}s_{13}e^{i\delta_{\rm CP}} & c_{23}c_{13}
\end{array}\right),
\end{equation}
where $s_{ij}=\sin\theta_{ij}$ and $c_{ij}=\cos\theta_{ij}$.

Regardless of the success of the oscillation phenomenon, there is still room for new physics affecting neutrino phenomenology. In particular, several works have studied non-standard interactions~\cite{GonzalezGarcia:1998hj,Bergmann:2000gp,Guzzo:2004ue,Gago:2001si,Gago:2001xg,Fogli:2007tx,Ohlsson:2012kf,Esmaili:2013fva}, decoherence in oscillations~\cite{Gago:2000qc,Gago:2002na,Fogli:2003th,Morgan:2004vv,Lisi:2000zt,Hooper:2004xr,Farzan:2008zv,Oliveira:2010zzd,robertoPhd,Oliveira:2013nua,Berryman:2014yoa} and fast neutrino decay~\cite{Frieman:1987as,Raghavan:1987uh,Berezhiani:1991vk,Berezhiani:1992ry,Berezhiani:1993iy,Barger:1999bg,Beacom:2002cb,Joshipura:2002fb,Bandyopadhyay:2002qg,Ando:2004qe,Fogli:2004gy,PalomaresRuiz:2005vf,GonzalezGarcia:2008ru,Maltoni:2008jr,Baerwald:2012kc,Meloni:2006gv,Das:2010sd,Dorame:2013lka,Gomes:2014yua,Berryman:2014qha,Berryman:2014yoa,Picoreti:2015ika,Abrahao:2015rba,Bustamante:2016ciw}, among other new physics effects~\cite{Pantaleone:1992ha,Bustamante:2010nq,Barger:2000iv,Colladay:1998fq, Esmaili:2014ota}.

In the last years, the interplay between oscillations and fast neutrino decay has been studied in a model-independent manner~\cite{Lindner:2001fx,PalomaresRuiz:2005vf}. This has been achieved by considering that the decay products of the neutrino decay are invisible to the detector. Such a situation can happen, for example, if the decay product consists of lighter, sterile states~\cite{Barger:1999bg,Beacom:2002cb,Joshipura:2002fb,Bandyopadhyay:2002qg,Ando:2004qe,Fogli:2004gy,GonzalezGarcia:2008ru,Maltoni:2008jr,Baerwald:2012kc,Meloni:2006gv,Das:2010sd,Dorame:2013lka,Gomes:2014yua,Berryman:2014qha,Berryman:2014yoa,Picoreti:2015ika,Abrahao:2015rba}. Throughout this work, we refer to this scenario as {\it invisible decay}. In this situation, one can bound the neutrino lifetime-to-mass ratio $\tau/m$ or, alternatively, the parameter $\alpha=E\,\Gamma$, related to the neutrino width $\Gamma$ evaluated at the energy $E$. In terms of these parameters, the two most important constraints to date are the following:
\begin{itemize}
 
 \item For solar neutrinos, the studies in~\cite{Berryman:2014qha,Picoreti:2015ika} have constrained the lifetime of $\nu_2$, giving $\tau_2/m_2\geq7.2\times10^{-4}~{\rm s}\cdot{\rm eV}^{-1}$ at $99\%$ C.L., equivalent to $\alpha_2<9.1\times10^{-13}$ eV$^2$.

 \item In~\cite{GonzalezGarcia:2008ru}, invisible decay was taken into account within atmospheric neutrino experiments. The authors combined this data with old data from MINOS~\cite{Weber:2008zzc}, bounding the $\nu_3$ lifetime. The study gave $\tau_3/m_3>3.0\times10^{-10}~{\rm s}\cdot{\rm eV}^{-1}$ at $90\%$ C.L., equivalent to $\alpha_3<2.2\times10^{-6}$ eV$^2$.
\end{itemize}

Although not as competitive as~\cite{GonzalezGarcia:2008ru}, one should also take into account the work in~\cite{Gomes:2014yua}. Here, MINOS and T2K $\nu_\mu$ disappearance data was used to constrain invisible decay, giving a limit of $\tau_3/m_3\geq2.8\times10^{-12}~{\rm s}\cdot{\rm eV}^{-1}$, at $90\%$ C.L.. This corresponds to $\alpha_3<2.4\times10^{-4}$~eV$^2$.

In contrast, not much work has been done in the direction of {\it visible decay}, where the final decay products involve active neutrinos, which can be detected. In oscillation experiments, this manifests itself as an additional contribution to invisible decay. We refer to the addition of both visible and invisible contributions as {\it full decay}. Nevertheless, in order to describe the visible contribution, one needs an expression for the differential width, so it is not possible to carry out this analysis in a completely model-independent way.

Models for fast neutrino decay usually involve the coupling of two neutrinos with a massless scalar, called a Majoron~\cite{Chikashige:1980qk,Gelmini:1983ea,Schechter:1981cv,Dias:2005jm}. This interaction can proceed through scalar ($g_s$) or pseudoscalar ($g_p$) couplings. As we shall see in this paper, both couplings can be related to the $\alpha$ decay constant by $g^2_{s,p}\sim16\pi\alpha/m^2$, where $m$ is the neutrino mass. It is then of our interest to understand how such an interaction affects the results of oscillation experiments, for both possible kinds of coupling.

To this end, we implement a full decay formalism within two experiments: MINOS and T2K. As mentioned before, the $\nu_\mu\to\nu_\mu$ disappearance channels of both experiments were used in~\cite{Gomes:2014yua} to study invisible decay. Of course, one of our objectives is to update the results in these channels within the full decay framework. However, this time we also have data from the T2K $\nu_\mu\to\nu_e$ appearance channel, for both neutrinos and antineutrinos. As far as we are aware of, this is the first time that visible decay is considered as modification of a neutrino appearance phenomenon, and we find it has important consequences.

Before we proceed, we need to point out that important constraints on neutrino - Majoron couplings exist~\cite{Kachelriess:2000qc,Pasquini:2015fjv,Gando:2012pj,Agostini:2015nwa,Hannestad:2005ex,Archidiacono:2013dua}. If one assumes that these can be directly translated as bounds on the decay constant $\alpha$, and assuming for illustration a neutrino mass $m=0.01$~eV, they would imply constraints as low as $\alpha<\ord{10^{-25}}$~eV$^2$. However, this translation is not always straightforward, and most of the bounds rely on additional assumptions, apart from the Majoron hypothesis. Thus, constraints coming from neutrino decay within oscillation experiments play an important role in transparently understanding the kind of phenomenology these new couplings bring, as they are established under controlled experimental conditions. Therefore, in the following, in addition to updating the bounds in~\cite{Gomes:2014yua}, we give a detailed explanation of how the neutrino spectrum of long-baseline experiments is modified, hoping that these insights might be useful in other scenarios.

We present our framework in Section~\ref{sec:osc}, and provide insight on how the neutrino transition operator will be modified by the different kind of coupling. In Section~\ref{sec:setup}, we describe our experimental setup for T2K and MINOS, as well as describe our procedure when carrying out the fit. In Section~\ref{sec:results} we present our results for each experiment, and then combine the data. Since the combination does not provide a better fit in comparison to the one from standard oscillations, we extract bounds on our decay parameter $\alpha$. We conclude in Section~\ref{sec:conclusions}.

We also include three Appendices. In Appendix~\ref{app:formulae.kin}, we present the general, model independent, full decay framework, within the oscillation scenario. In Appendix~\ref{app:formulae.dyn} we give details of the neutrino - Majoron coupling, and explain how we obtain the formulae used in this work. Finally, in Appendix~\ref{app:expformulae} we describe how we carry out the MINOS and T2K simulations.

\section{Neutrino Visible Decay in Oscillation Experiments}
\label{sec:osc}

We consider a neutrino oscillation experiment, where a neutrino flux is directed towards a detector located a large distance away. During their propagation, the neutrinos are subject to an evolution function which takes into account their oscillation and decay, such that the flux arriving at the detector is modified.

To calculate the flux of $\nu_\beta$ and $\bar\nu_\beta$ arriving at the detector, with energy $E_\beta$, due to the oscillation and decay of unstable $\nu_\alpha$ and $\bar\nu_\alpha$ ($\alpha,\beta=e,\mu$), we use:
\begin{equation}
 \label{eq:basic}
 \frac{dN^s_\beta}{dE_\beta}=\sum_{\alpha,r}\int dE_\alpha\,P_{\alpha \beta}^{rs}(E_\alpha,\,E_\beta)\,\phi_{\nu_\alpha^r}(E_\alpha)~.
\end{equation}
Here, $\phi_{\nu_\alpha^r}(E_\alpha)$ is the original flux of $\nu^r_\alpha$ with energy $E_{\alpha}$, and $P_{\alpha\beta}^{rs}(E_\alpha,\,E_\beta)$ describes the neutrino oscillation and decay process to $\nu^s_\beta$. The $r,s=\{(+),\,(-)\}$ indices refer to the neutrino chirality (i.e. $\nu^{(-)}$ for $\nu$, and $\nu^{(+)}$ for $\bar\nu$). The evolution function $P_{\alpha\beta}^{rs}(E_\alpha,\,E_\beta)$ is obtained by adapting a relevant formula in~\cite{PalomaresRuiz:2005vf}:
\begin{eqnarray}
\label{eq:oscdecay}
P_{\alpha\beta}^{rs}(E_\alpha,\,E_\beta) &=& 
\left|\sum_{i}U_{\beta i}^{(s)}U_{\alpha i}^{(r)*}\exp\left[-i\frac{m_i^2L}{2E_\alpha}\right]\exp\left[-\frac{\alpha_iL}{2E_\alpha}\right]\right|^2\delta(E_\alpha-E_\beta)\delta_{rs} \nonumber \\
&&+G_{\alpha\beta}^{rs}(E_\alpha,\,E_\beta)~,
\end{eqnarray}
where $U^{(-)}=U_{\rm PMNS}$ and $U^{(+)}=U^*_{\rm PMNS}$. The Lorentz-invariant variable $\alpha_i$ is related to the total neutrino decay width $\Gamma_i$, such that $\alpha_i=E_\alpha\,\Gamma_i$.

The first term in the evolution function includes the standard oscillation contributions, multiplied by an exponential governed by the neutrino decay parameter $\alpha_i$. For stable neutrino mass eigenstates, all $\alpha_i$ are zero, and we recover the standard oscillation formula. In the following, we shall refer to this term as the contribution from {\it invisible decay} (ID).

If we included this first term only, we would be describing the decay of the mass eigenstates into lighter non-interacting states. However, we contemplate the possibility that these lighter states consist of active neutrinos with lower energy. This is taken into account by the second term, which we refer to as the contribution from {\it visible decay} (VD). For the rest of this work, we refer to the addition of the two terms as {\it full decay} (FD).

The general formula for the visible decay function $G_{\alpha\beta}^{rs}(E_\alpha,\,E_\beta)$ can be found on Appendix~\ref{app:formulae.kin}. In this work we use a simplified version, considering only one decay channel ($\nu^r_i\to\nu^s_f J$, where $\nu^r_i$ and $\nu^s_f$ are on the mass basis). It has the following form:
 \begin{equation}
  \label{eq:nudecay1}
 G_{\alpha\beta}^{rs}(E_\alpha,\,E_\beta)= 
  \left(1-e^{-\Gamma_i L}\right)\left|U^{(r)}_{\alpha i}\right|^2\left|U^{(s)}_{\beta f}\right|^2 
  \frac{1}{\Gamma_i}\frac{d}{dE_\beta}\Gamma(\nu_i^r\to\nu_f^s\,J)~.
 \end{equation}
We can understand Eq.~(\ref{eq:nudecay1}) in the following way. The term $(1-e^{-\Gamma_i L})\left|U_{\alpha i}\right|^2$, when multiplied by the original flux $\phi_{\nu_\alpha^r}$, gives the amount of $\nu^r_i$ mass eigenstates that decay. When the latter is multiplied by the normalized spectrum, $\tfrac{1}{\Gamma_i}\tfrac{d}{dE_\beta}\Gamma(\nu_i^r\to\nu_f^s\,J)$, we obtain the number of $\nu_f^s$ mass eigenstates, with energy between $E_\beta$ and $E_\beta+d E_\beta$, produced from the decay. The amount of final $\nu_\beta^s$ flavour eigenstates is then obtained by changing back to the flavour basis with $\left|U_{\beta f}\right|^2$. 

The full expression for $d\Gamma(\nu_i^r\to\nu_f^s\,J)/dE_\beta$ is derived from the interaction Lagrangian, which is written in terms of scalar ($g_s$) or pseudoscalar ($g_p$) neutrino - Majoron couplings. Details are given in Appendix~\ref{app:formulae.dyn}. Furthermore, if we take only one non-vanishing coupling (either $g_s$ or $g_p$), the final expression is greatly simplified:
\begin{equation}
\label{eq:simpleG}
G_{\alpha\beta}^{rs}(E_\alpha,\,E_\beta)= \left|U^{(r)}_{\alpha i}\right|^2\left|U^{(s)}_{\beta f}\right|^2
 \left(\frac{1-e^{-\alpha_i L/E_\alpha}}{E_\alpha}\right)F^{\prime rs}_g(E_\alpha,\,E_\beta)~,
\end{equation}
with the visible decay function given by
\begin{equation}
\label{eq:visdecfunc}
F^{\prime rs}_g(E_\alpha,\,E_\beta) = 
 \frac{x_{if}^2}{(x_{if}^2-1)}F^{rs}_g(E_\alpha,\,E_\beta)\times\Theta_H(E_\alpha-E_\beta)\,\Theta_H(x_{if}^2E_\beta-E_\alpha)~.
\end{equation}
Here we have $x_{if}=m_i/m_f>1$, the label $g=\{g_s,\,g_p\}$ indicates the non-vanishing coupling, $\Theta_H(x)$ is the Heaviside function and we have replaced the total width $\Gamma_i\to \alpha_i/E_{\alpha}$. The $F^{rs}_g(E_\alpha,\,E_\beta)$ functions have chirality conserving ($r=s$) and chirality changing ($r\neq s$) cases:
\begin{subequations}
\label{eq:BigF}
\begin{align}
 F^{\pm\pm}_{g_s}(E_\alpha,\,E_\beta) &= \frac{1}{E_\alpha\,E_\beta}\frac{(E_\alpha+x_{if}E_\beta)^2}{(x_{if}+1)^2}~,  &
 F^{\pm\mp}_{g_s}(E_\alpha,\,E_\beta) &= \frac{(E_\alpha-E_\beta)}{E_\alpha\,E_\beta}\frac{(x_{if}^2E_\beta-E_\alpha)}{(x_{if}+1)^2}~,  \\
 F^{\pm\pm}_{g_p}(E_\alpha,\,E_\beta) &= \frac{1}{E_\alpha\,E_\beta}\frac{(E_\alpha-x_{if}E_\beta)^2}{(x_{if}-1)^2}~, &
 F^{\pm\mp}_{g_p}(E_\alpha,\,E_\beta) &= \frac{(E_\alpha-E_\beta)}{E_\alpha\,E_\beta}\frac{(x_{if}^2E_\beta-E_\alpha)}{(x_{if}-1)^2}~.
\end{align}
\end{subequations}
For a given $x_{ij}$ and energy, we have that $\pm\pm$ and $\pm\mp$ transitions are complementary, which is expected, since $F_g^{\pm\pm}+F_g^{\pm\mp}=1$.

\begin{figure}[tbp]
\centering
\includegraphics[width=0.45\textwidth]{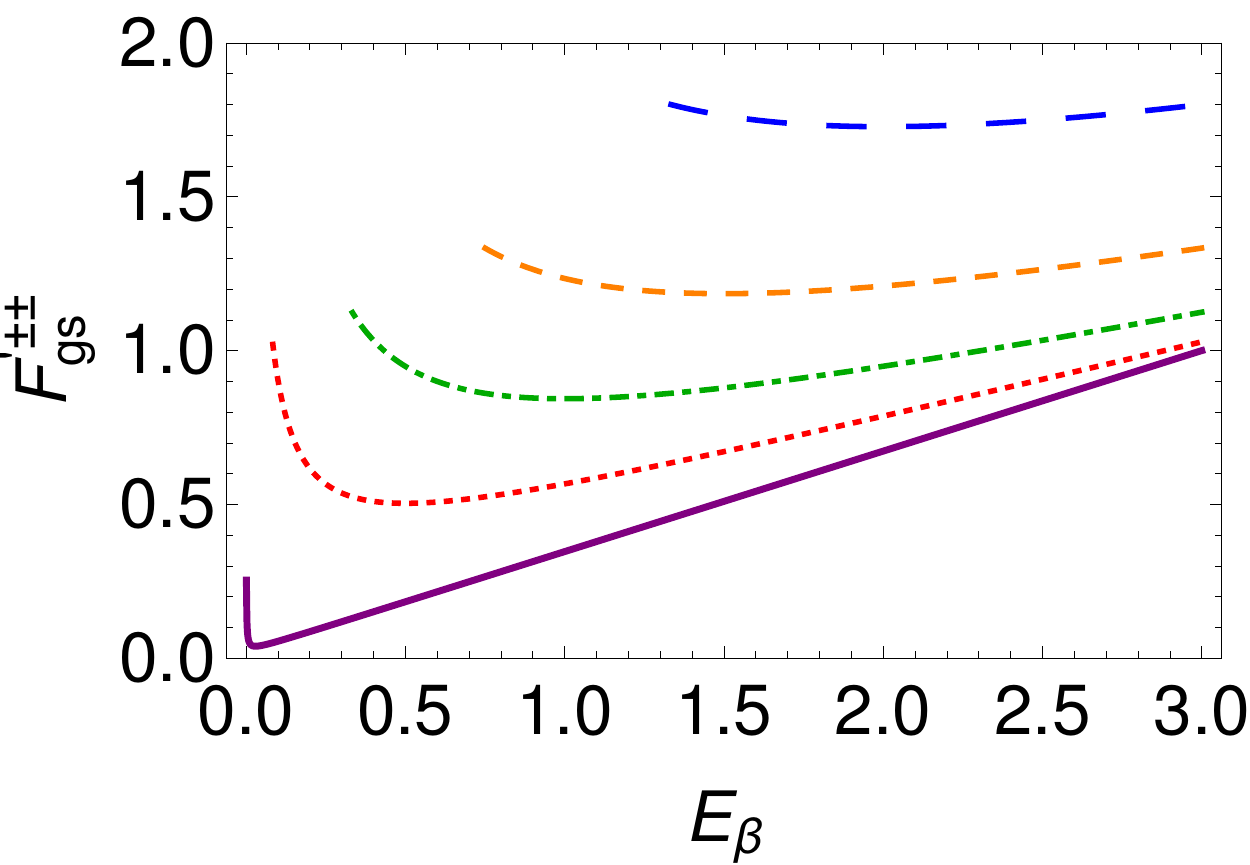} \hfill
\includegraphics[width=0.45\textwidth]{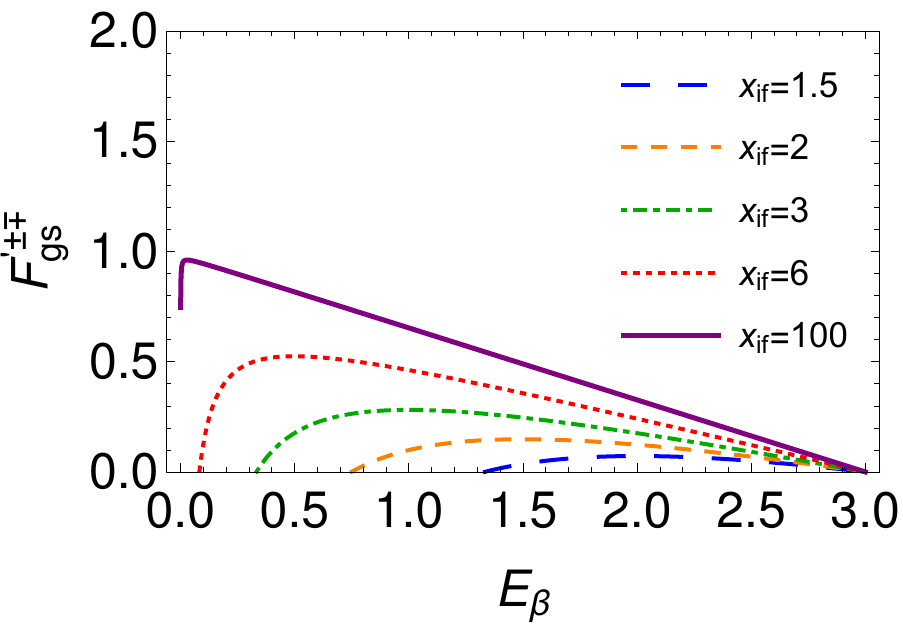} \\
\includegraphics[width=0.45\textwidth]{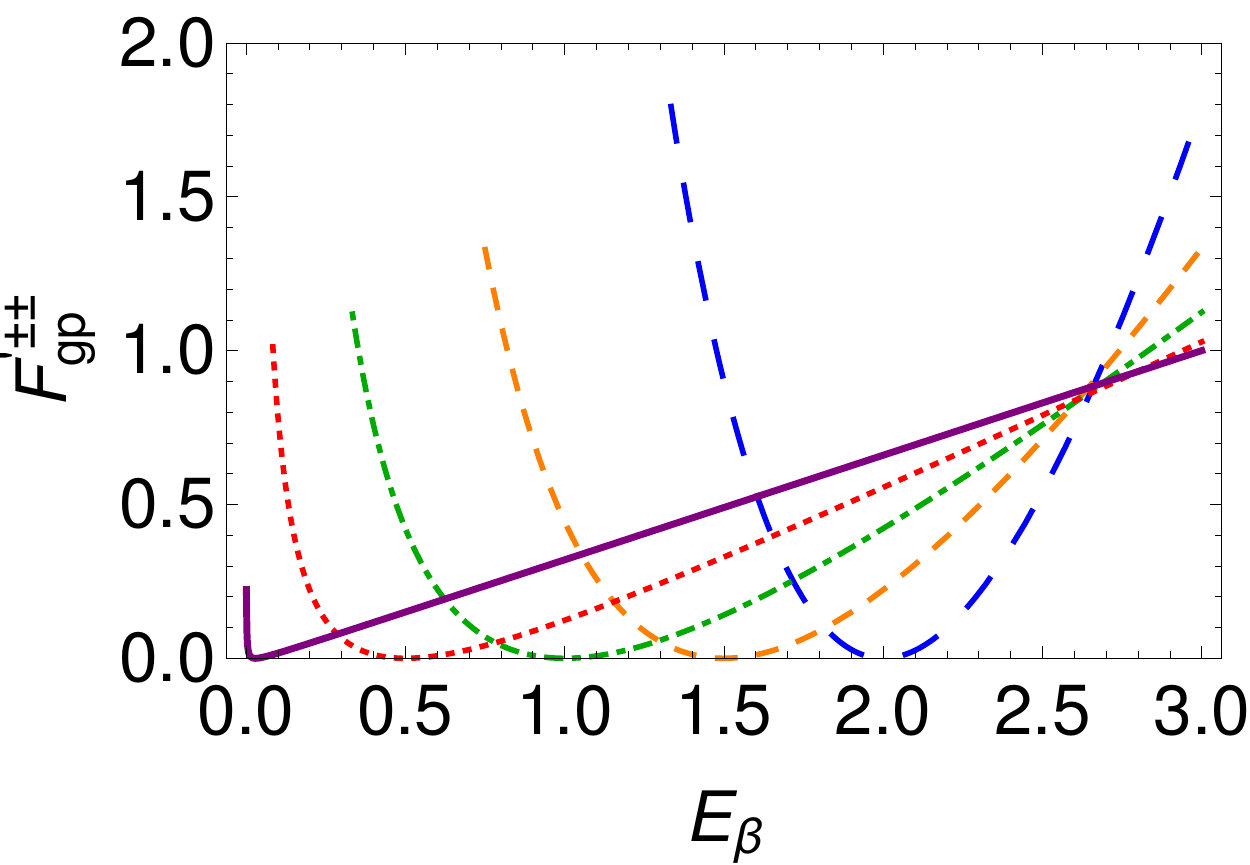} \hfill
\includegraphics[width=0.45\textwidth]{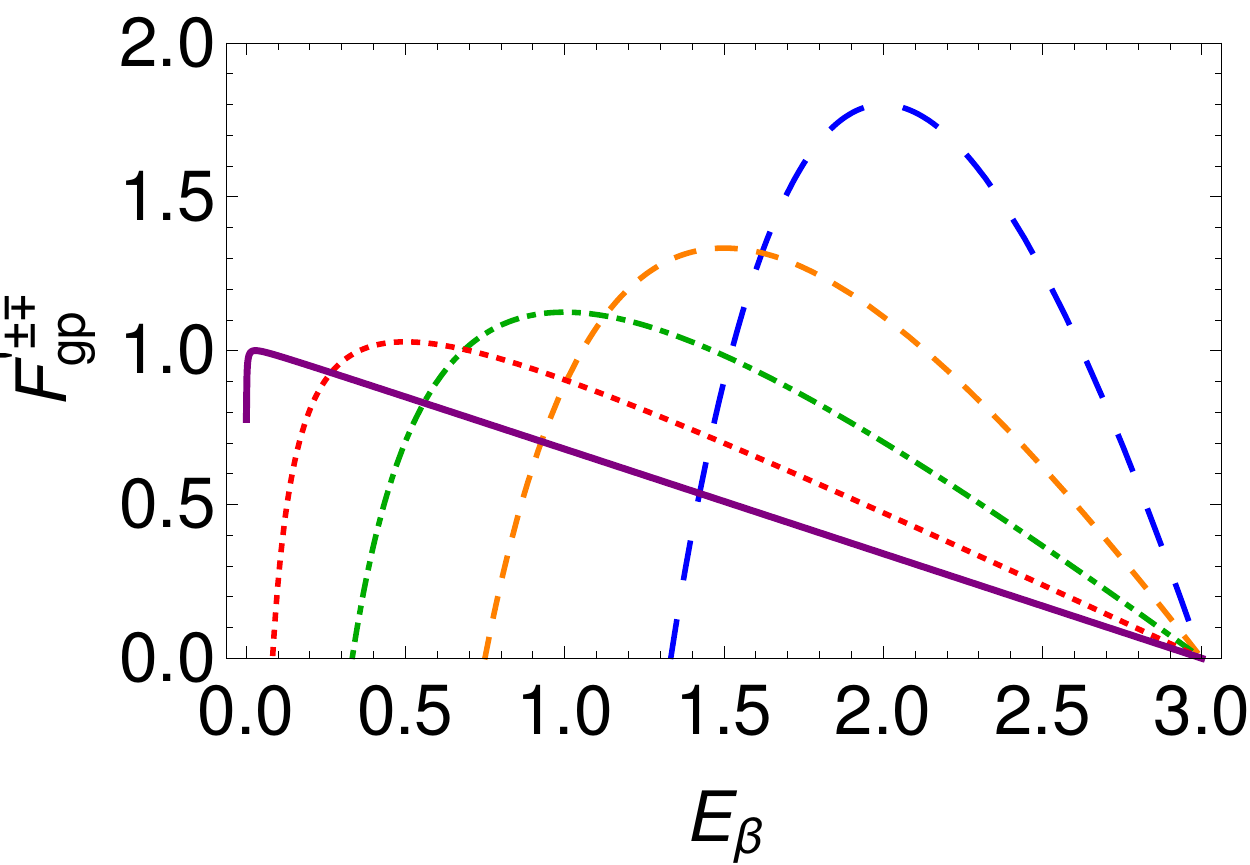}
\caption{Visible decay functions $F^{\prime rs}_g$, as functions of final energy $E_\beta$ and $x_{if}$. The top (bottom) row shows the function for a scalar (pseudoscalar) coupling, while the left (right) column shows the ones for chirality conserving (changing) processes. We have fixed the initial neutrino energy as $E_\alpha=3$~GeV.}
\label{fig:Ffuncs}
\end{figure}

Studying the behaviour of these functions is important in order to understand how scalar and pseudoscalar couplings affect our analysis of T2K and MINOS. To this end, we show in Figure~\ref{fig:Ffuncs} the functions $F^{\prime rs}_g$, defined in Eq.~(\ref{eq:visdecfunc}), for fixed $E_\alpha=3$ GeV, and for several values of $x_{if}$, as a function of $E_{\beta}$. On the top (bottom) rows we show results for scalar (pseudoscalar) couplings. The left column shows chirality-conserving transitions ($\pm\pm$), while the right columns shows chirality-changing transitions ($\pm\mp$).

For large $x_{if}$ (purple curve), we see that both scalar and pseudoscalar couplings have the same behaviour (compare purple curve within the same column).  For $E_\beta$ close to $E_\alpha$, we find that $\pm\pm$ processes (left column) are favoured, while for lower energies, $\pm\mp$ processes (right column) dominate.

For lower $x_{if}$ (dashed blue curve) we have a different behaviour depending on the coupling. For scalar couplings, we see that $\pm\pm$ transitions dominate all final energies. Meanwhile, pseudoscalar couplings have a mixed behaviour, with a clear preference for $\pm\mp$ transitions. 

From now on, we assume a normal ordering scenario, with $\nu_3$ unstable and decaying exclusively into $\nu_1$. We label $\alpha_3\to\alpha$, $m_1\to m_{\rm light}$ and $(g_{s,p})_{31}\to g_{s,p}$ (see Appendix~\ref{app:formulae.dyn}). Notice that cosmological bounds~\cite{Ade:2013zuv} require $\sum_i m_i<0.23 $~eV. For normal ordering, and taking squared mass differences at their best fit points, for example, this would rule out $x_{31}<1.24$. This bound is taken into account in our final result.

\section{Experimental Setup and Fitting Procedure}
\label{sec:setup}

\subsection{T2K}

The T2K experiment~\cite{Abe:2015awa} has been running in the latest years, sending a neutrino beam to the Super-Kamiokande (SK) detector~\cite{Kajita:2016vhj}, located 295 km away from the source. The detection process uses Cerenkov radiation to identify neutrinos, however, this is blind to the charge of the associated lepton. Thus, it is unable to distinguish between neutrinos and antineutrinos, as a consequence, in this work we shall sum the neutrino and antineutrino contributions when calculating the number of events. The experiment has two running modes, called ``neutrino'' and ``antineutrino'' runs.

The neutrino run of the T2K experiment consisted in delivering a primarily $\nu_\mu^{(-)}$ beam to the SK detector, with a final luminosity of $7.48\times10^{20}$ protons on target (POT). Their results in the $\nu^{(-)}_\mu\to\nu_\mu^{(-)}$ disappearance channel imposed strong bounds on the $\sin^22\theta_{23}$ and $\Delta m^2_{32}$ parameter space~\cite{Abe:2012gx,Abe:2015awa}. In addition, through the observation of $\nu_\mu^{(-)}\to\nu_e^{(-)}$ appearance channel, it provided the first indication of non-zero $\theta_{13}$~\cite{Abe:2011sj}.

In the current antineutrino run, a primarily $\nu_\mu^{(+)}$ beam illuminates the SK detector. The luminosity released to the public corresponds to $7.47\times10^{20}$ POT, and the combination of both runs has given hints favouring a negative value of $\delta_{\rm CP}$~\cite{ICHEP-T2K-2016,NOW-T2K-2016}. In our analysis we use both datasets.

On the following, we shall focus on the T2K neutrino run. For the antineutrino run, the same analysis applies, taking the CP conjugates of the states.

Given that $\nu_\mu^{(-)}$ is the largest component in the beam~\cite{Abe:2015awa}, we use the FD mode (visible and invisible contributions), for both appearance and disappearance channels. For the other components of the flux ($\nu_\mu^{(+)}$, $\nu_e^{(-)}$ and $\nu_e^{(+)}$), we only include the FD when the visible part is not negligible.

For the $\nu_\mu$ disappearance channel, we include FD for $\nu_\mu^{(-)}\to\nu_\mu^{(\pm)}$ and ID for $\nu_\mu^{(+)}\to\nu_\mu^{(+)}$. As usual, the signal consists of charged-current quasielastic (CCQE) interactions, while the main background includes the charged-current non-quasielastic (CCnQE) and neutral current (NC) interactions. 

The $\nu_e$ appearance channel has many contributions. As mentioned previously, we consider the FD for $\nu^{(-)}_\mu\to\nu^{(\pm)}_e$. In this case, we also include the FD for $\nu^{(+)}_\mu\to\nu^{(\pm)}_e$, but take only the ID contribution for $\nu^{(\pm)}_e\to\nu^{(\pm)}_e$. Both CCQE and CCnQE interactions are considered part of the signal. The background consists of the $\nu^{(\pm)}_e\to\nu^{(\pm)}_e$ processes, as well as NC contributions.

We also include full decay when considering the NC contribution to the events. Notice that the latter shall now depend on the neutrino mixing angles, due to the non-unitarity of the decay mechanism~\cite{Berryman:2014yoa}.

We give more details on the generation of events at T2K in Appendix~\ref{app:t2kformulae}.

\subsection{MINOS}

MINOS was a long-baseline neutrino experiment \cite{Adamson:2013whj} which used two detectors and was exposed to a neutrino beam produced at Fermilab (NuMI beam line~\cite{Adamson:2015dkw}). The latter is a two-horn-focused neutrino beam that can be configured in two ways: Forward Horn Current (FHC), to produce a beam optimized for muon neutrinos, and Reverse Horn Current (RHC), for a beam optimized for muon anti-neutrinos. The Near and the Far Detectors are located at $1$ km and $735$ km from the target, respectively.

The data set used in our analysis comes from FHC mode with $10.71\times10^{20}$ POT. The beam composition was $92.9\%$ of $\nu_\mu^{(-)}$, $5.8\%$ of $\nu_\mu^{(+)}$, and $1.3\%$ of $\nu_e^{(\pm)}$. We use the data set that comprises the charged current (CC) contained-vertex neutrino and anti-neutrino disappearance data \cite{Adamson:2013whj, Cao:2014eca}. 

Notice that, in contrast to T2K, the MINOS magnetized muon spectrometer does distinguish between neutrinos and antineutrinos. Thus, even though we only use the FHC mode, we do include the $\nu_\mu^{(-)}$ and $\nu_\mu^{(+)}$ beam components in our analysis, separately.

In order to probe the decay framework against the $\nu_\mu^{(-)}$ disappearance data we consider the FD for the $\nu_\mu^{(-)} \rightarrow \nu_\mu^{(-)}$ channel and only the VD $\nu_\mu^{(+)}\rightarrow \nu_\mu^{(-)}$ contribution. Analogously, for $\nu_\mu^{(+)}$ disappearance data, we take into account the FD for $\nu_\mu^{(+)} \rightarrow \nu_\mu^{(+)}$ and the VD $\nu_\mu^{(-)}\rightarrow \nu_\mu^{(+)}$.

The details about the MINOS reconstruction data can be found in Appendix~\ref{app:minosformulae}.

\subsection{Statistical Analysis for T2K and MINOS}
\label{sec:statistics}

The relevant parameters in this study, which we shall vary, are $s^2_{23}$, $s^2_{13}$, $\delta_{\rm CP}$, $\Delta m^2_{32}$, $\alpha$ and $m_{\rm light}$. We keep fixed $s^2_{12}=0.306$ and $\Delta m^2_{21}=7.5\times10^{-5}$~eV$^2$.

To perform the fit for T2K, we use a $\chi^2$ function similar to the one used in~\cite{Abe:2015awa}. In order to take into account systematic errors, we include normalisation and energy calibration nuisance parameters, $n_x$ and $t_x$, respectively, for signal ($x=s$) and background ($x=b$). For a given set of oscillation and decay parameters, and for each channel, the $\chi^2$ is minimized with respect to $n_x$ and $t_x$, adding appropriate pull factors~\cite{Huber:2002mx,Fogli:2002pt}:
\begin{multline}
 \label{t2kchi}
 \chi^2(s^2_{23},\,s^2_{13},\,\delta_{\rm CP},\,\Delta m^2_{32},\,\alpha,\,m_{\rm light})=\\\sum_{\beta=e,\mu}\,\min_{\{n_x,\,t_x\}}\left(2\sum_{\rm bins}\left(N_{\beta,\,\rm fit}-N_{\beta,\,\rm obs}\log N_{\beta,\,\rm fit}\right)+\left\{\frac{n_s^2}{\sigma^2_{n s}}+\frac{n_b^2}{\sigma^2_{n b}}+\frac{t_s^2}{\sigma^2_{t s}}+\frac{t_b^2}{\sigma^2_{t b}}\right\}_\beta\right)~.
\end{multline}
Here, $N_{\beta,\,\rm fit}$ is the sum of expected signal and background events per bin, which also involve the nuisance parameters. Denoting the numerical prediction per bin for signal and background $\nu_\beta^{(\pm)}$ events as $N^{(\pm)}_{\beta,\,s}$ and $N^{(\pm)}_{\beta,\,b}$, respectively, we have
\begin{equation}
 N_{\beta,\,\rm fit}=\sum_{\substack{x=s,b\\ r=+,-}}\left(1+n_x+t_x\frac{E_{\rm bin}-\hat E}{E_{\rm max}-E_{\rm min}}\right)_\beta N^r_{\beta,\,x}~,
\end{equation}
where $E_{\rm min}$ ($E_{\rm max}$) is the minimum (maximum) energy in the analyzed neutrino spectrum, $E_{\rm bin}$ is the average bin energy, and $\hat E=\tfrac{1}{2}(E_{\rm max}+E_{\rm min})$ is the average energy of the spectrum. 

The other parameters appearing in Eq.~(\ref{t2kchi}) are $N_{\beta,\,\rm obs}$, which is the observed number of $\nu_\beta^{(-)}$ and $\nu_\beta^{(+)}$ events~\cite{ICHEP-T2K-2016,NOW-T2K-2016}, and $\sigma_{n x}$ and $\sigma_{t x}$ are the respective uncertainties in normalisation and tilt, set both equal to $10\%$.

In the case of MINOS, using a similar notation, we use the following $\chi^2$ function:
\begin{equation}
\chi^2 = \sum_{\substack{{\rm bins}\\r=+,-}}\left(\frac{(1+n_s) N^r_{\mu,\,s} + (1+n_b) N^r_{\mu,\,b} - N_{\mu,\,\rm obs}}{\sigma_{\rm bin}}\right)^2 +  \frac{n_s^2}{\sigma^2_{n s}} + \dfrac{n_b^2}{\sigma^2_{n b}}~. 
\end{equation} 
The parameter $\sigma_{\rm bin}$ represents the statistical and systematic uncertainty extracted from data set, while $\sigma_{n s}$, $\sigma_{n b}$ are equal to $14.7\%$ and $4\%$, respectively~\cite{Adamson:2007gu}.

Finally, we can also include information from reactor data. For example, the Daya Bay experiment~\cite{An:2015nua} has given the most precise bound on $s^2_{13}$, which is $\sin^22\theta_{13}=0.092\pm0.017$~\cite{An:2012eh}. We have estimated the impact of neutrino decay on this experiment, and find that the ratio of the background-subtracted spectra to prediction assuming no oscillation is modified within the given error bars~\cite{An:2016ses}. We then assume that no reactor disappearance experiment is affected by neutrino decay. Thus, only for our final result constraining the decay parameter $\alpha$, we include an additional pull factor on our $\chi^2$ function:
\begin{equation}
 \chi^2_{\rm reactor}=\left(\frac{s_{13}^2-s_{13,\rm reactor}^2}{\sigma_{13,\rm reactor}}\right)^2~,
\end{equation}
with $s_{13,\rm reactor}^2=0.0243$ and $\sigma_{13,\rm reactor}=0.0026$, following the same procedure as in~\cite{Abe:2015awa}.

In our fit, we vary $m_{\rm light}$ and $\alpha$ logarithmically, from $5\times10^{-3}$ to $10^{-1}$~eV and $2\times10^{-6}$ to $5\times10^{-4}$~eV$^2$, respectively. The oscillation angles $\theta_{13}$ and $\theta_{23}$ are also scanned, with $0\leq s_{13}^2\leq0.05$ and $0.32\leq s_{23}^2\leq0.86$. The phase $\delta_{\rm CP}$ is scanned between $-\pi$ and $\pi$, while the squared mass difference $\Delta m^2_{32}$ is varied between $1.6\times10^{-3}$ and $2.8\times10^{-3}$~eV$^2$.

\section{Impact on Oscillation Parameter Regions}
\label{sec:results}

\subsection{Analysis in T2K}
\label{sec:constraints.T2K}

In the following, we explore how the inclusion of the neutrino FD distorts the currently allowed regions for oscillation parameters, obtained from standard oscillations (SO).

\begin{figure}[!t]
\centering
\includegraphics[width=0.45\textwidth]{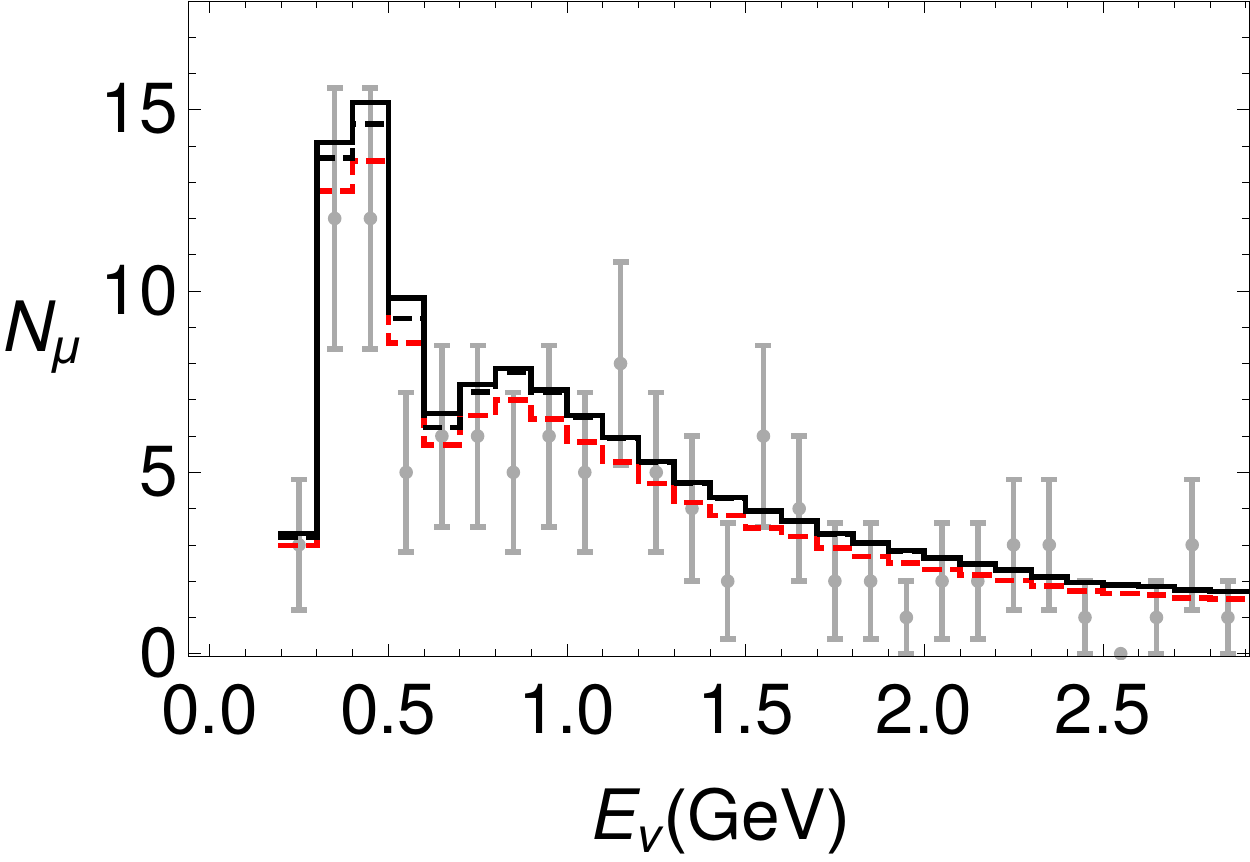} \hfill
\includegraphics[width=0.45\textwidth]{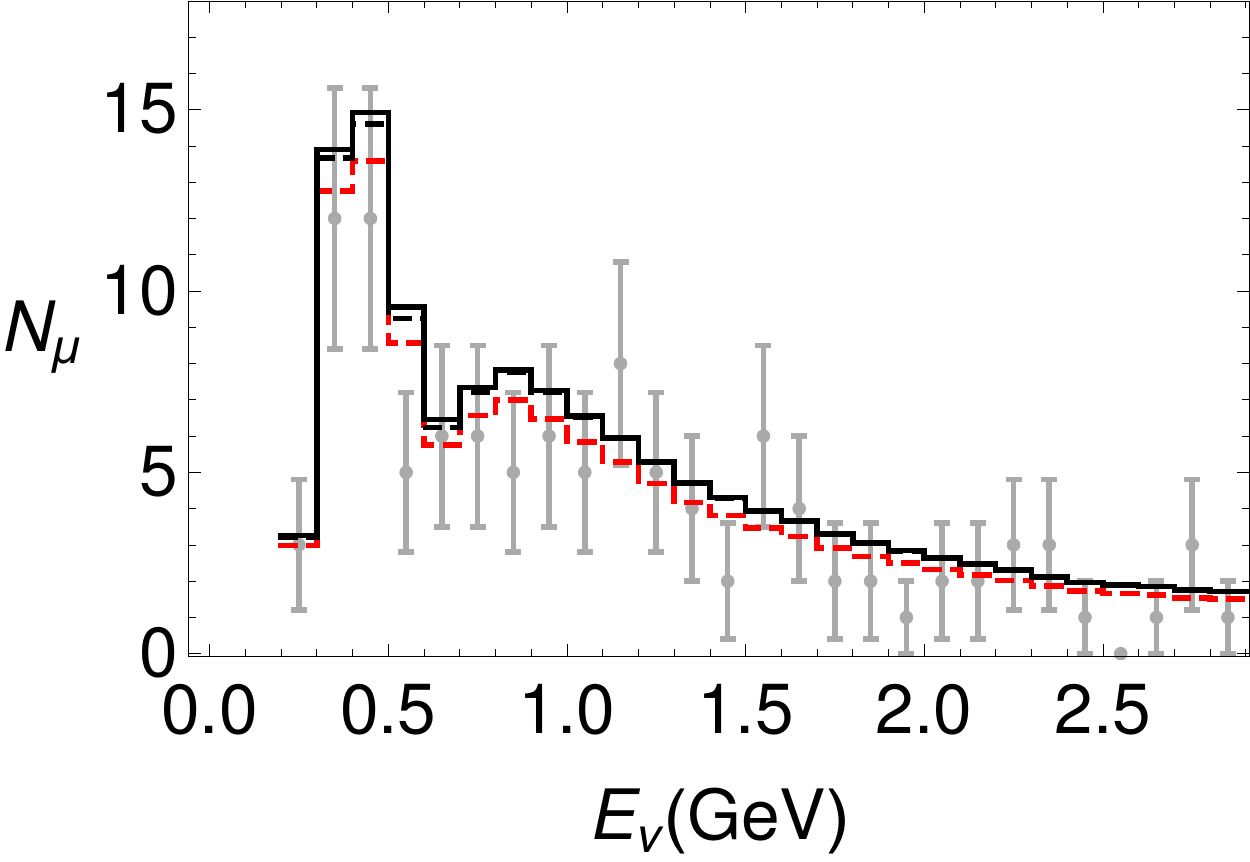} \\
\includegraphics[width=0.45\textwidth]{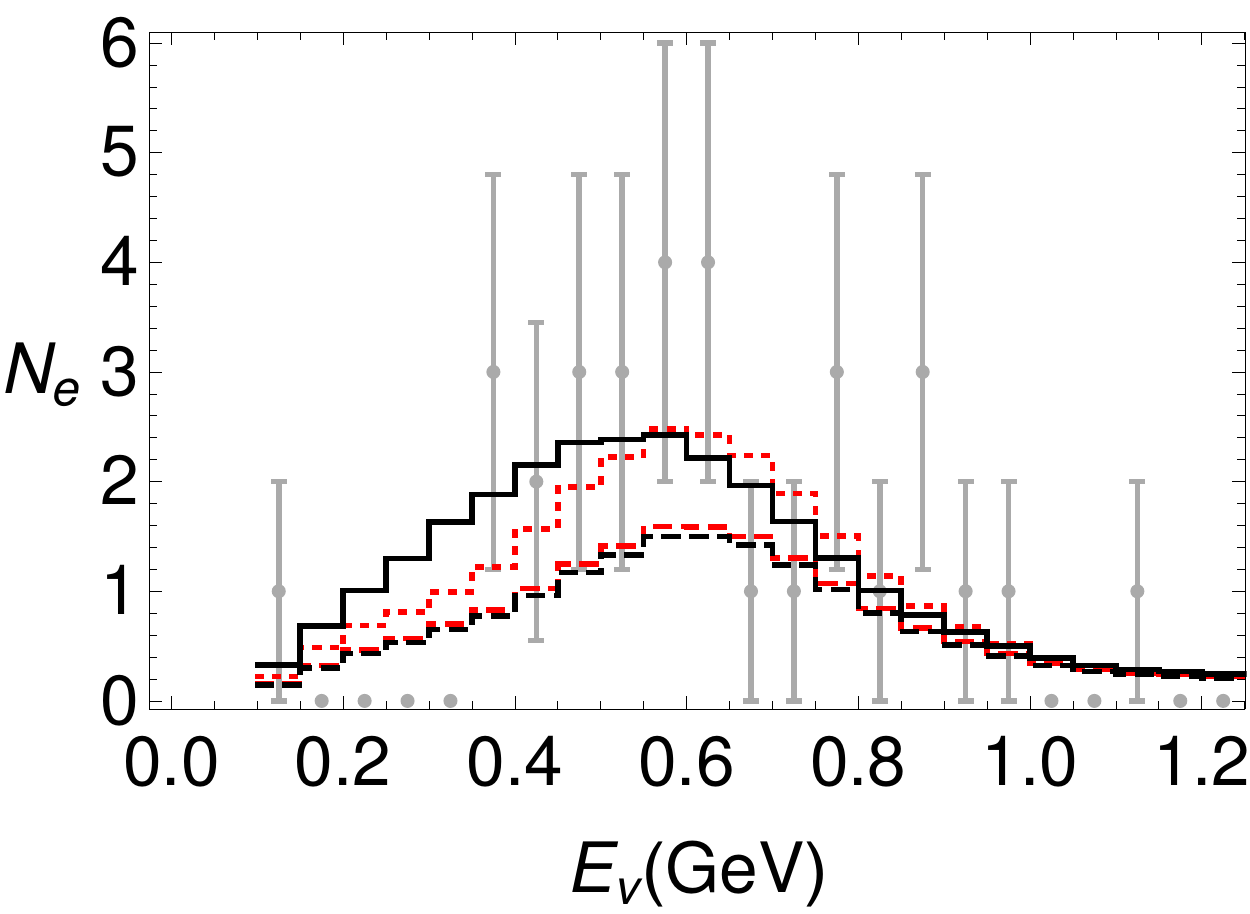} \hfill
\includegraphics[width=0.45\textwidth]{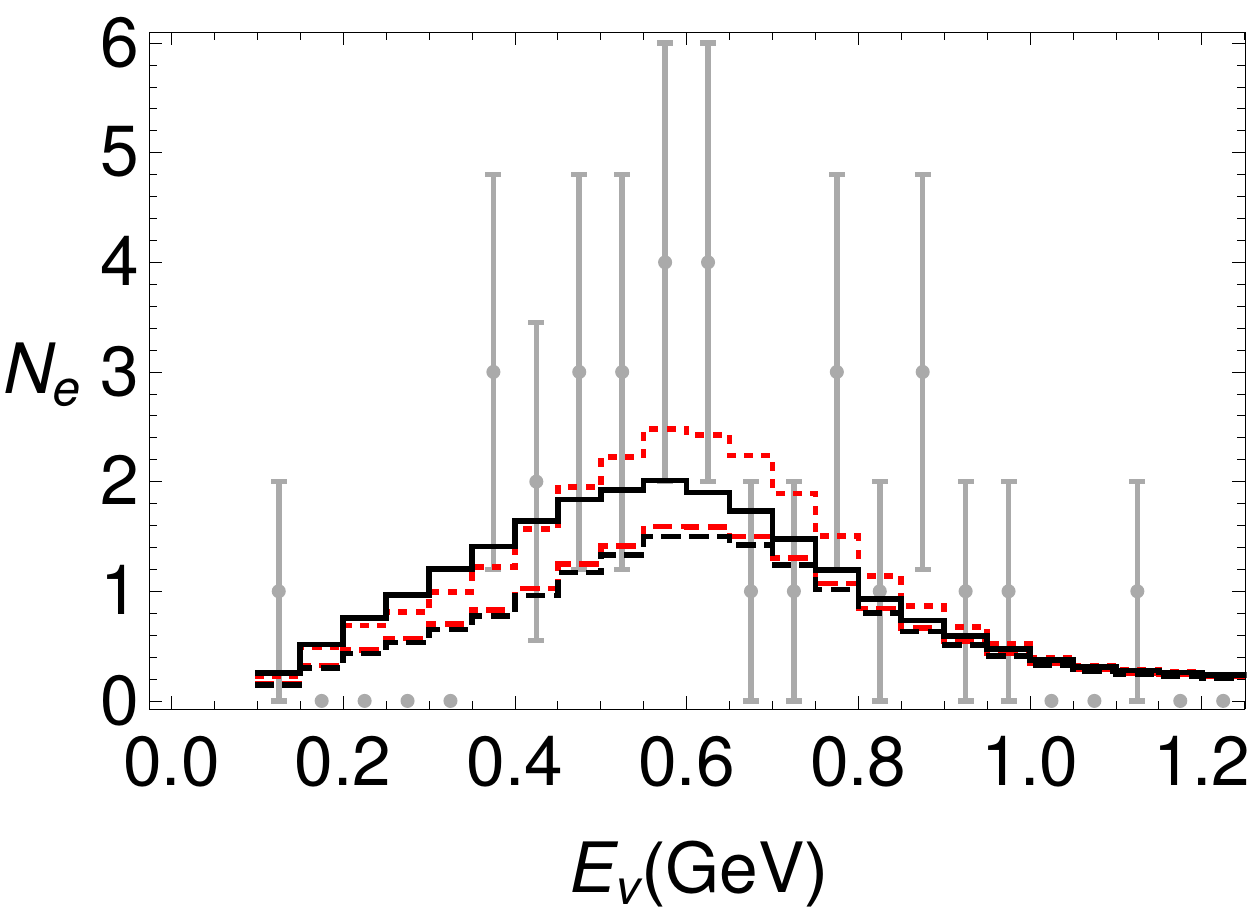}
\caption{Spectrum of disappearance (top row) and appearance (bottom row) events, for the neutrino run. The spectrum for SO is shown in red, dotted (dashed) for $\delta_{\rm CP}=-\pi/2$ ($+\pi/2$). The spectrum for ID is shown dashed, in black, and the spectrum for FD is shown solid, in black. For both we set $\delta_{\rm CP}=+\pi/2$. We show results for scalar (pseudoscalar) couplings on the left (right). T2K data is shown in gray~\cite{ICHEP-T2K-2016}.}
\label{fig:spectrum.nu}
\end{figure}
In Figure~\ref{fig:spectrum.nu}, we show the spectrum of neutrino appearance and disappearance events for the neutrino run, for SO, ID and FD. We have fixed $s^2_{23}=0.532$, $s^2_{13}=0.022$ and $\Delta m^2_{32}=2.545\times10^{-3}$~eV$^2$, which correspond to the best-fit parameters for T2K data. We have set $\alpha=4\times10^{-5}$~eV$^2$, which is roughly $10\%$ of the mean lifetime that T2K should be sensitive to ($\approx E/L$). We have also set $m_{\rm light}=7\times10^{-2}$ eV to maximize the difference between the scalar and pseudoscalar scenarios (see Figure~\ref{fig:Ffuncs}).

The top panels show neutrino disappearance events. We see that for the chosen value of $\alpha$, there is no big difference between the SO and FD scenarios. Nevertheless, even though not shown, one finds that for larger values of $\alpha$ the decay scenario erases the oscillation dip. This is due to neutrino decay behaving as a decoherence term in the oscillation formulae.

The bottom panels show neutrino appearance events. Here, the SO scenario is shown for $\delta_{\rm CP}=\pm\pi/2$. Notice that one obtains a larger amount of events when $\delta_{\rm CP}=-\pi/2$. In contrast, we show ID and FD scenarios only with $\delta_{\rm CP}=+\pi/2$. A first important feature is that, although the ID contribution is similar to the SO result (for the same value of $\delta_{\rm CP}$), when comparing the FD and ID spectra it is evident that the VD contribution is sizeable. Thus, this decay scenario has a stronger impact on appearance than on disappearance measurements. A second important feature is that the FD scenario, with $\delta_{\rm CP}=\pi/2$, can have a similar number of events as the SO result with $\delta_{\rm CP}=-\pi/2$ at the cost of bringing some distortion to the low-energy (high-energy) part of the spectrum, for scalar (pseudoscalar) couplings.

\begin{figure}[!t]
\centering
\includegraphics[width=0.45\textwidth]{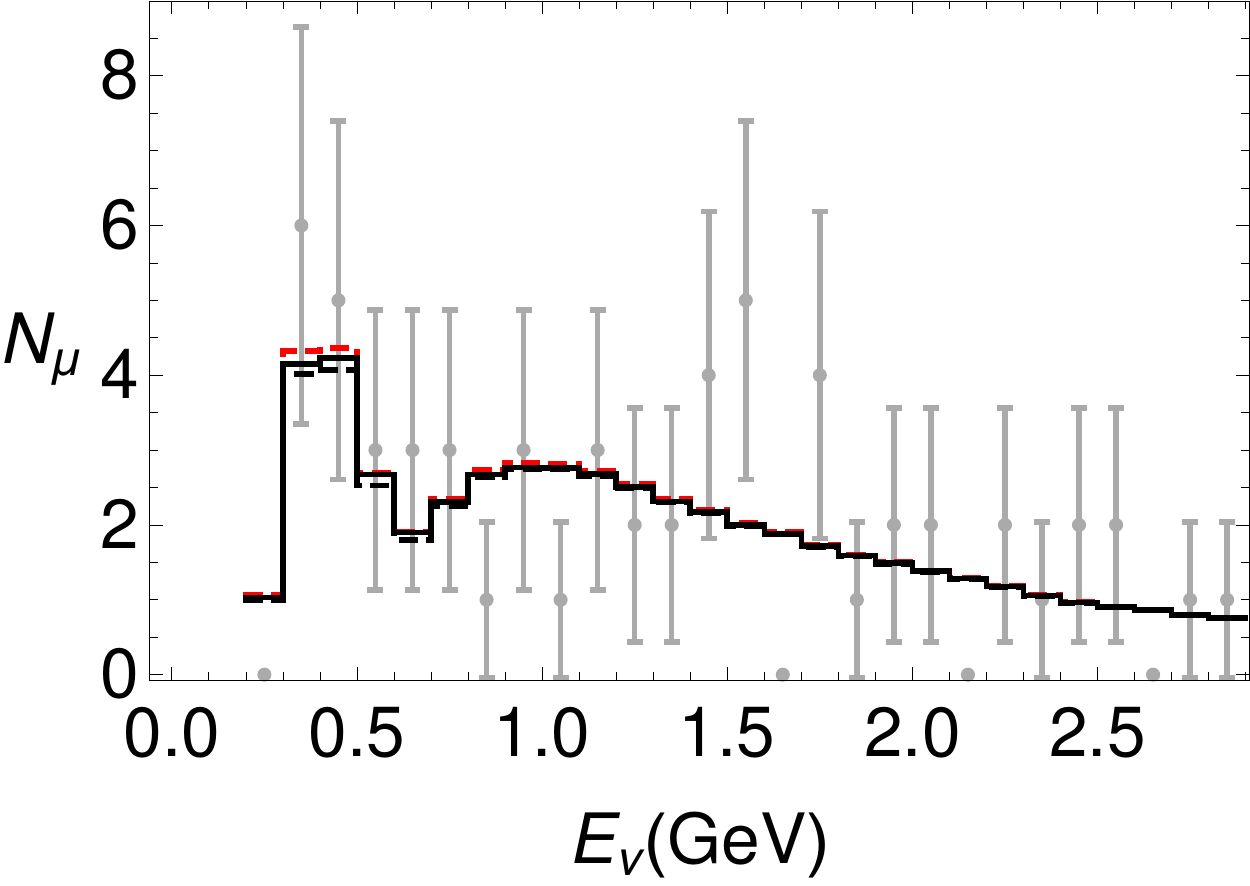} \hfill
\includegraphics[width=0.45\textwidth]{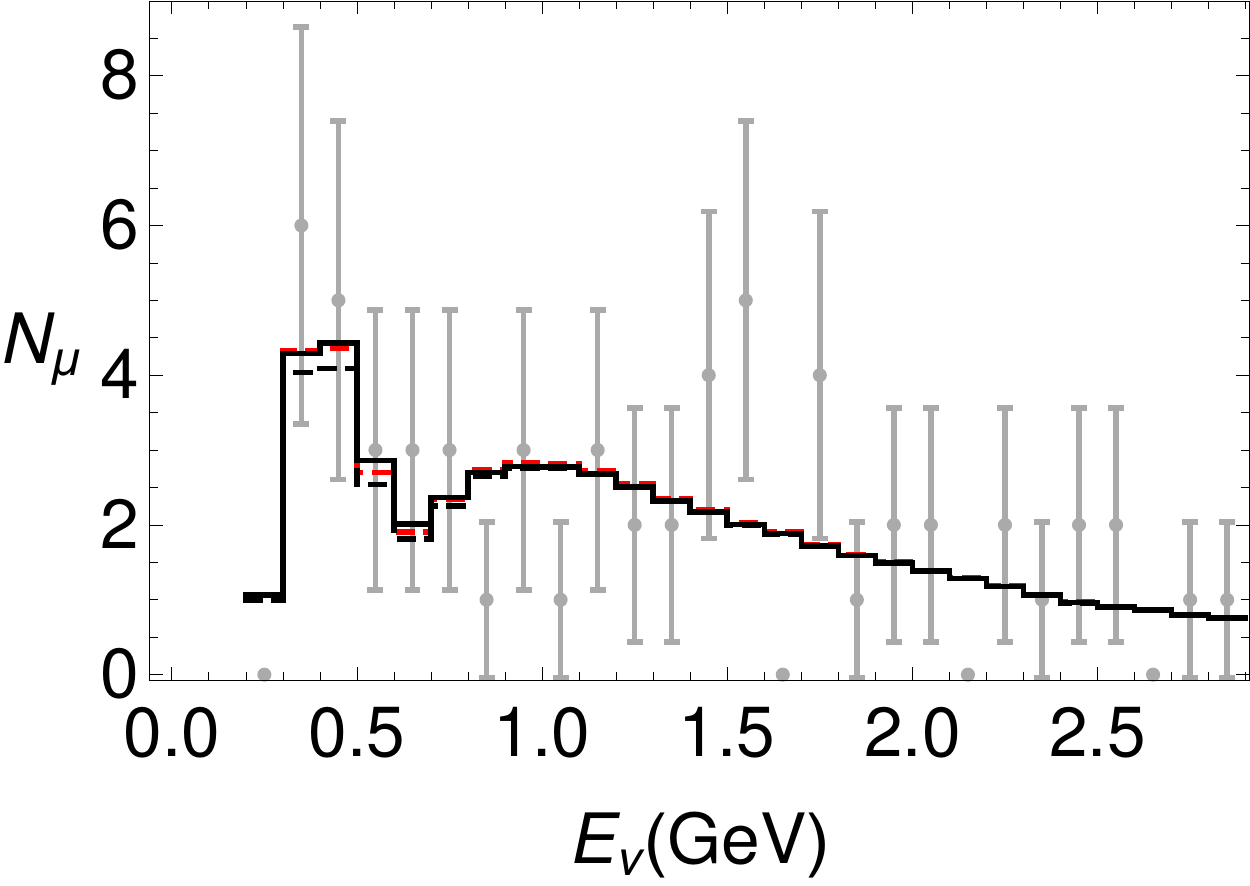} \\
\includegraphics[width=0.45\textwidth]{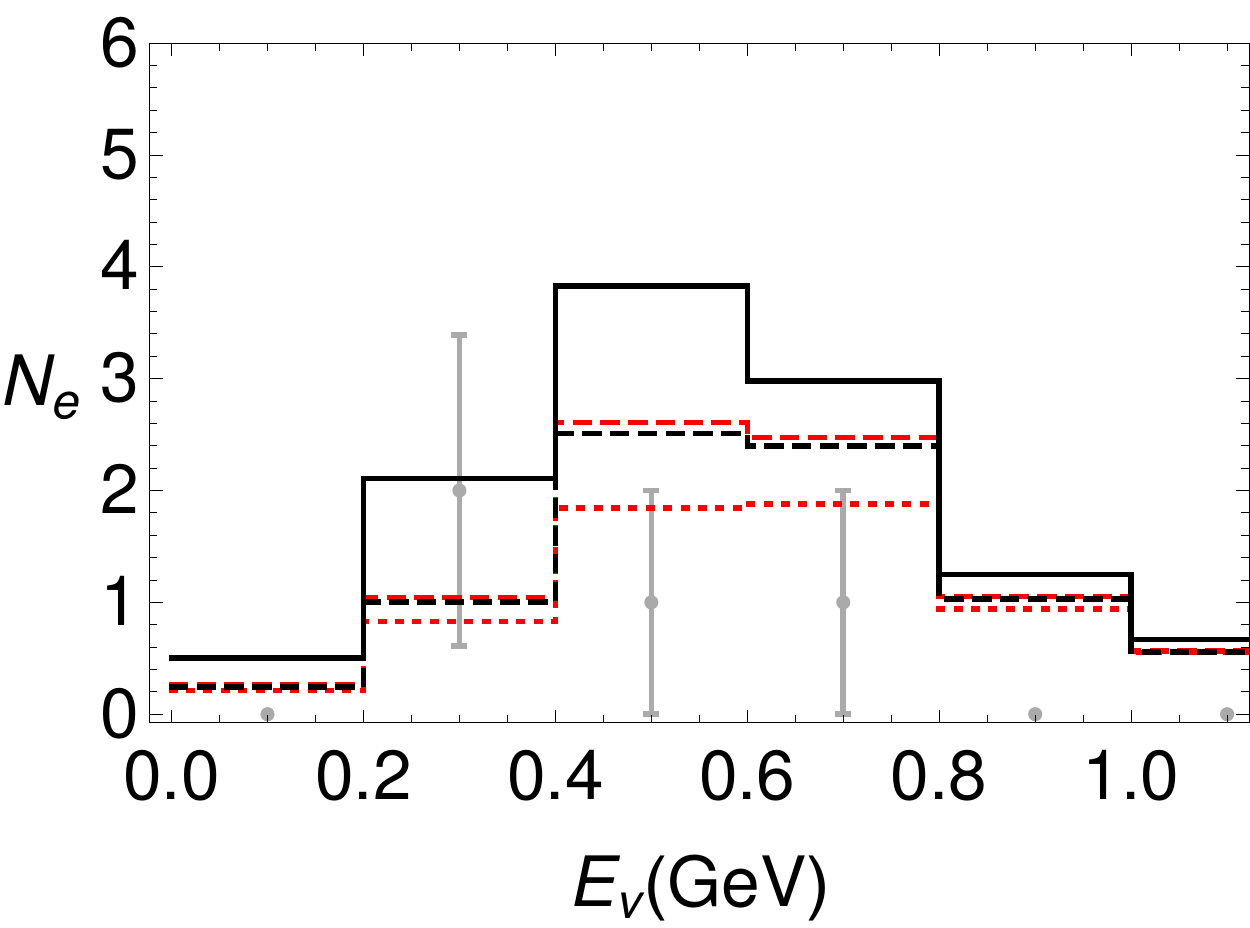} \hfill
\includegraphics[width=0.45\textwidth]{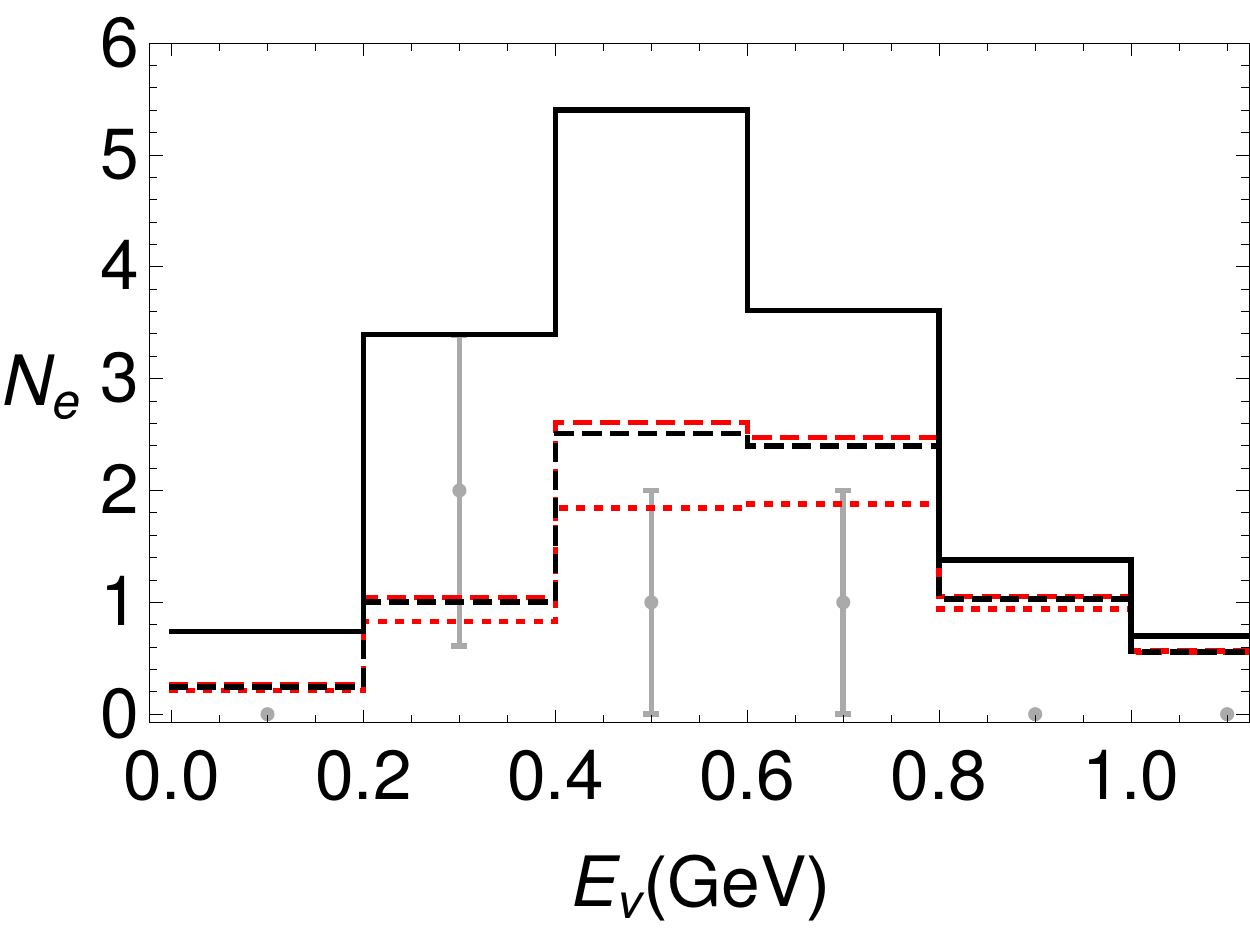}
\caption{As in Figure~\ref{fig:spectrum.nu}, but for the antineutrino run, with T2K data from~\cite{NOW-T2K-2016}.
\label{fig:spectrum.anu}}
\end{figure}
We show in Figure~\ref{fig:spectrum.anu} the corresponding spectra for the antineutrino run. For disappearance events (top row), the conclusions are similar to those for Figure~\ref{fig:spectrum.nu}, that is, for the given value of $\alpha$, neutrino decay does not significantly modify the spectrum.

On the other hand, as in Figure~\ref{fig:spectrum.nu}, for the appearance spectrum we show SO with $\delta_{\rm CP}=\pm\pi/2$. On this case, $\delta_{\rm CP}=-\pi/2$ leads to a smaller number of events compared to $\delta_{\rm CP}=+\pi/2$. We again show the spectrum for ID and FD with $\delta_{\rm CP}=+\pi/2$, and again find the VD contribution to be sizeable. However, contrary to the neutrino run, this time the number of events greatly exceeds the SO prediction for $\delta_{\rm CP}=-\pi/2$. In fact, the ID contribution alone is already too large. This means that the antineutrino run shall be relevant when disentangling the value of $\delta_{\rm CP}$ within the decay scenario.

On both Figures we find that the FD results differ for scalar and pseudoscalar couplings. The reason for this is that we have chosen a large $m_{\rm light}$. In the previous section (Figure~\ref{fig:Ffuncs}), we found that for scalar couplings, large values of $m_{\rm light}$ ($x_{31}\to1$) favour $\nu^{(\pm)}\to\nu^{(\pm)}$ over $\nu^{(\pm)}\to\nu^{(\mp)}$ transitions. The opposite behaviour is seen in pseudoscalar couplings, which allow a much larger proportion of chirality-changing transitions. This, of course, shall modify results, as $\nu^{(-)}$ and $\nu^{(+)}$ states have different cross-sections.

Given the argument above, we see that on the neutrino run the peak of the spectrum is smaller for the pseudoscalar than for the scalar coupling. As said previously, pseudoscalar couplings give a larger rate of $\nu^{(+)}$, and since these have a smaller cross-section, the number of events is lower. Furthermore, for the antineutrino run, the pseudoscalar coupling gives a larger rate of $\nu^{(-)}$, which have higher cross-sections, and thus lead to more events.

\begin{figure}[!t]
\centering
\includegraphics[width=0.45\textwidth]{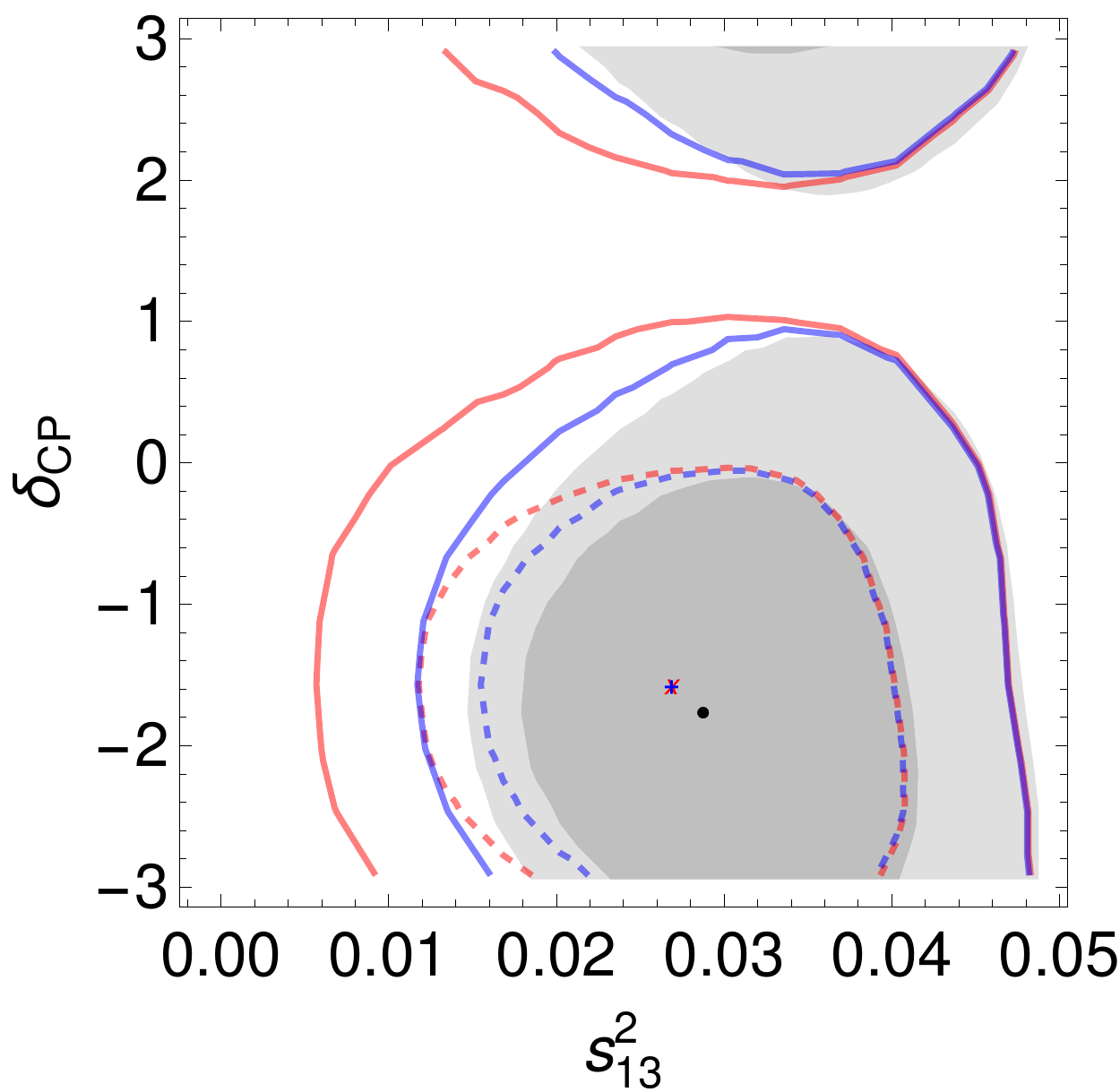} \hfill
\includegraphics[width=0.45\textwidth]{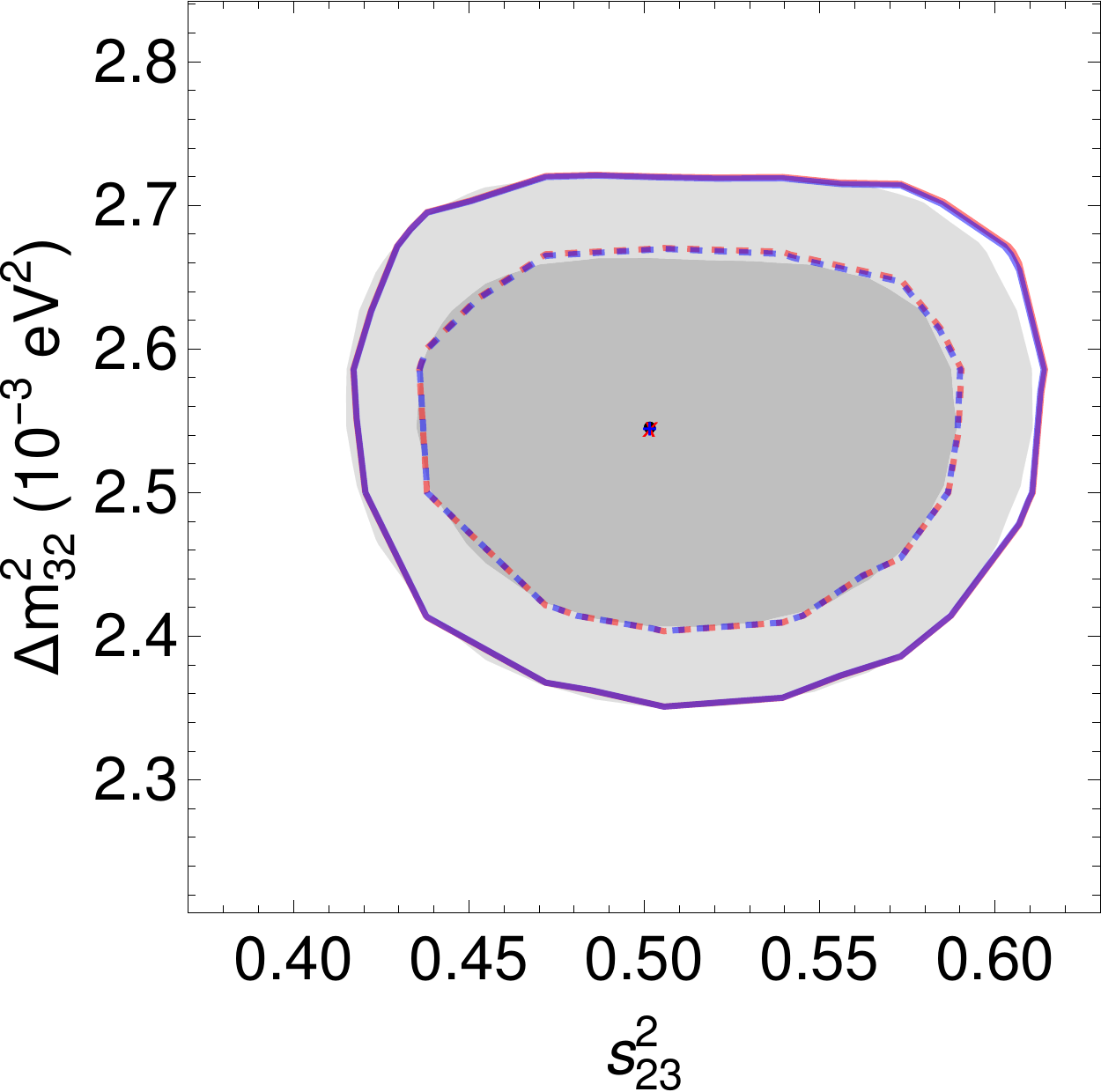} \\
\includegraphics[width=0.45\textwidth]{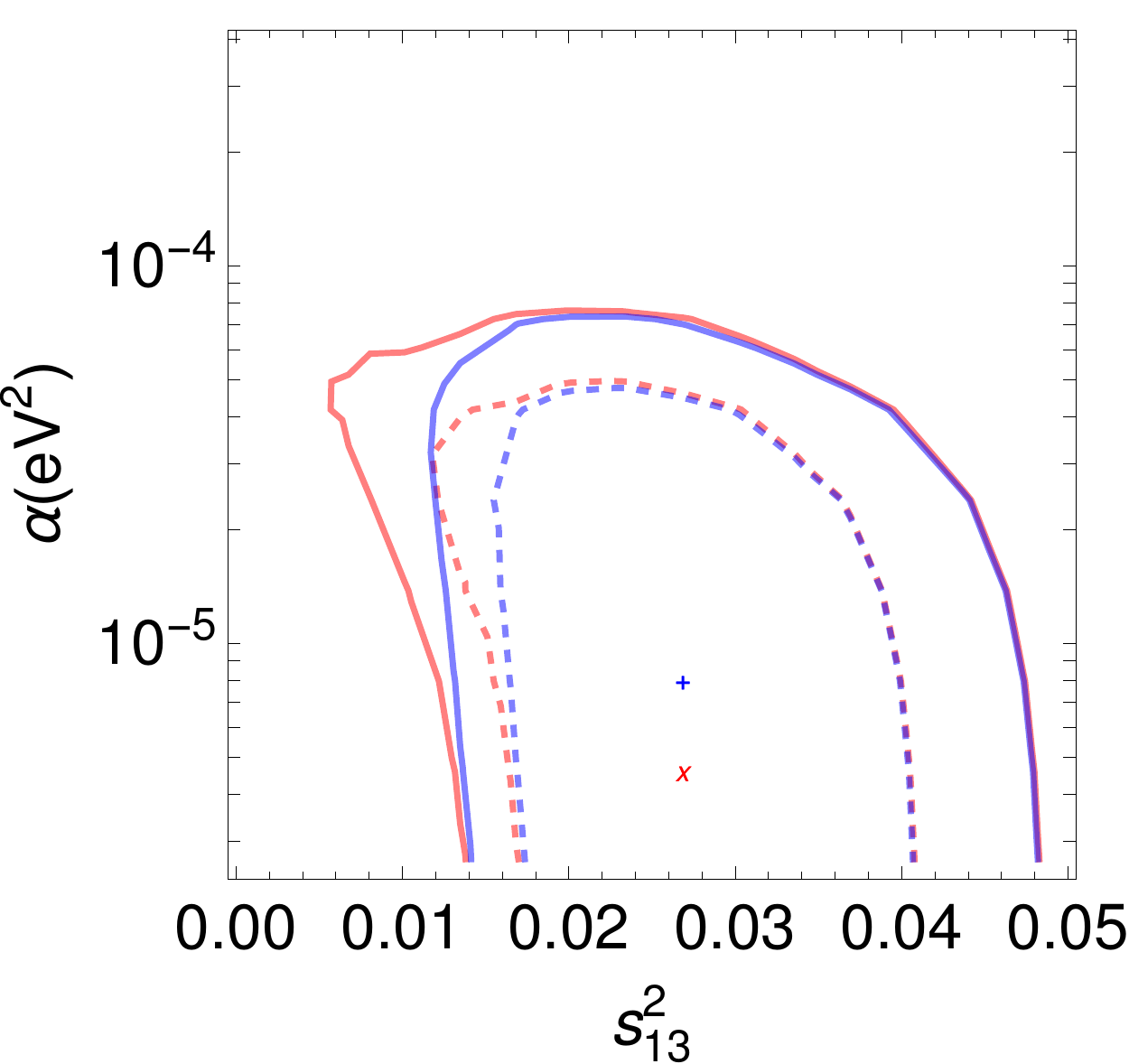} \hfill
\includegraphics[width=0.45\textwidth]{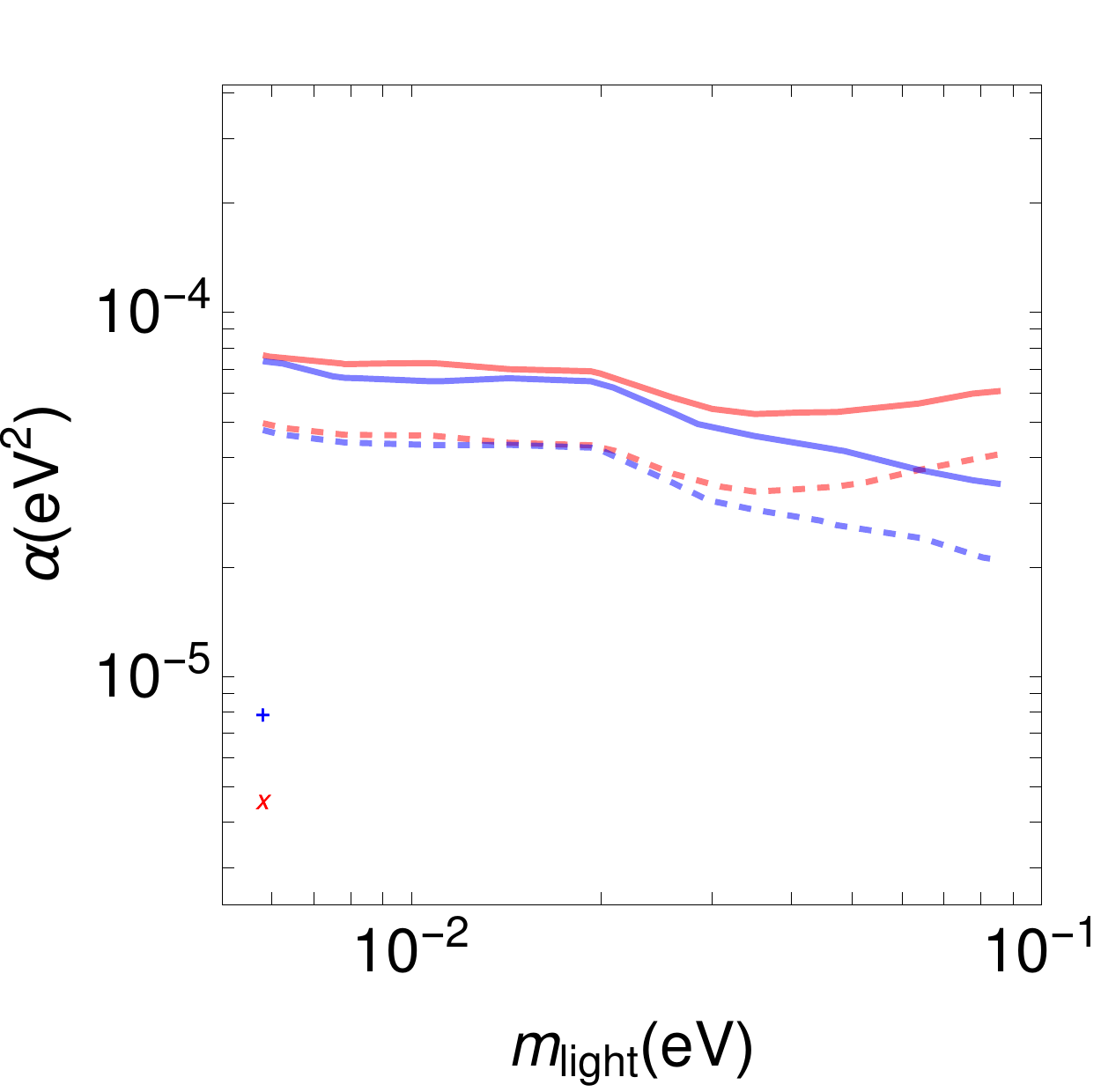}
\caption{Correlations in neutrino parameters, for scalar (red) and pseudoscalar (blue) couplings, using data from the T2K neutrino and antineutrino runs. The dashed lines show $1\sigma$ contours, while the solid line indicates the $90\%$ C.L., as is done in the original T2K analysis. Shaded regions show corresponding contours in the SO scenario. The black dot, red~$\times$ and blue~$+$ indicate the best-fit points.
\label{fig:constraints1.nunubar}}
\end{figure}
In Figure~\ref{fig:constraints1.nunubar}, we show allowed regions in several subspaces of the parameter space, under the hypothesis of the FD scenario. In all plots, red (blue) curves refer to scalar (pseudoscalar) couplings. Moreover, we show the corresponding regions on the SO scenario in shades of grey.

If we concentrate on the $s_{23}^2-\Delta m^2_{32}$ subspace (top right panel), dominated by disappearance measurements, we notice that the resulting allowed regions for scalar and pseudoscalar couplings are equal, and very similar to those for the SO scenario. This confirms that the FD scenario has little impact on neutrino disappearance events.

The FD scenario has a much larger impact in the $s_{13}^2-\delta_{\rm CP}$ region (top left panel), where appearance measurements are crucial. We find that the upper bound to $s^2_{13}$ is similar to the one for SO, for both scalar and pseudoscalar couplings. However, the lowest possible value of $s^2_{13}$ decreases. Moreover, we find that a large region of positive $\delta_{\rm CP}$ is ruled out, in agreement with the T2K fit for the SO scenario. 

One can get a better insight on the situation in appearance measurements from the other panels, where we show the allowed values of $\alpha$ as a correlation with $s^2_{13}$ and $m_{\rm light}$.

On the bottom left panel, we see that for low values of $\alpha$, the allowed range for $s^2_{13}$ is well bounded (consistent with SO). Then, as $\alpha$ increases, the FD formalism starts to influence, and the range for $s^2_{13}$ is shifted to smaller values. This means that the lower number of neutrinos from oscillations, due to the smaller $s_{13}^2$, is compensated by the extra neutrinos coming from decay. Finally, in both cases, too large values of $\alpha$ generate too many appearance events, which is incompatible with T2K data, restricting the allowed $\alpha$ to values of $\ord{10^{-5}}$~eV$^2$. This automatically leaves unaffected the $s_{23}^2-\Delta m^2_{32}$ 
subspace, which requires much larger values of $\alpha$ to modify significantly the spectrum.

On the other hand, on the bottom right panel, one finds the confidence level contours for $\alpha$ as a function of $m_{\rm light}$. For scalar (pseudoscalar) couplings, we find that for large values of $m_{\rm light}$, the contours reach higher (lower) values of $\alpha$. For pseudoscalar couplings, large mass values favour chirality-changing transitions, which gives tensions when applied to both neutrino and antineutrino runs, as seen in Figures~\ref{fig:spectrum.nu}-\ref{fig:spectrum.anu}. Thus, in this case, smaller values of $\alpha$ are compatible with data. Opposed to the latter case, for scalar couplings with large $m_{\rm light}$, chirality-changing transitions are minimized, so visible decay has a smaller impact, and larger values of $\alpha$ are allowed. In contrast, for smaller $m_{\rm light}$, both scalar and pseudoscalar curves tend to the same value of $\alpha$. This means that in this limit both couplings are undistinguishable, as expected from Figure~\ref{fig:Ffuncs} (for example, for $x_{31}=100$).

Notice that the best-fit points correspond to non-vanishing values of $\alpha$. As we shall see later, these points are not statistically significant. Thus, from the T2K-only $\chi^2$ analysis, we conclude that the FD solution does not improve the quality of the fit in a statistically significant way. Therefore, from T2K data we will later obtain constraints on $\alpha$ (see Figure~\ref{fig:alphachi}).

\subsection{Analysis in MINOS}

\begin{figure}[!t]
\centering
\includegraphics[width=0.45\textwidth]{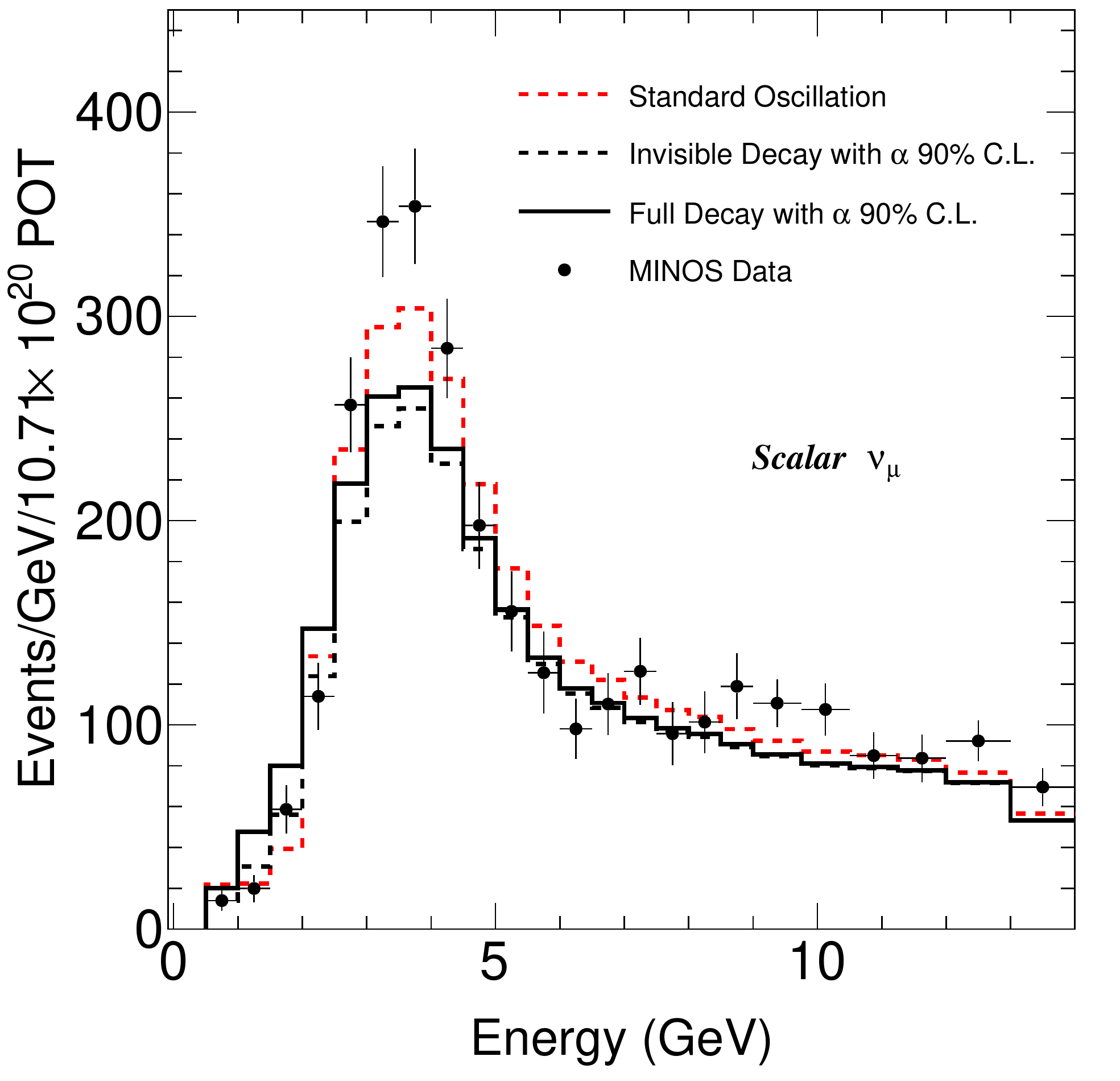} \hfill
\includegraphics[width=0.45\textwidth]{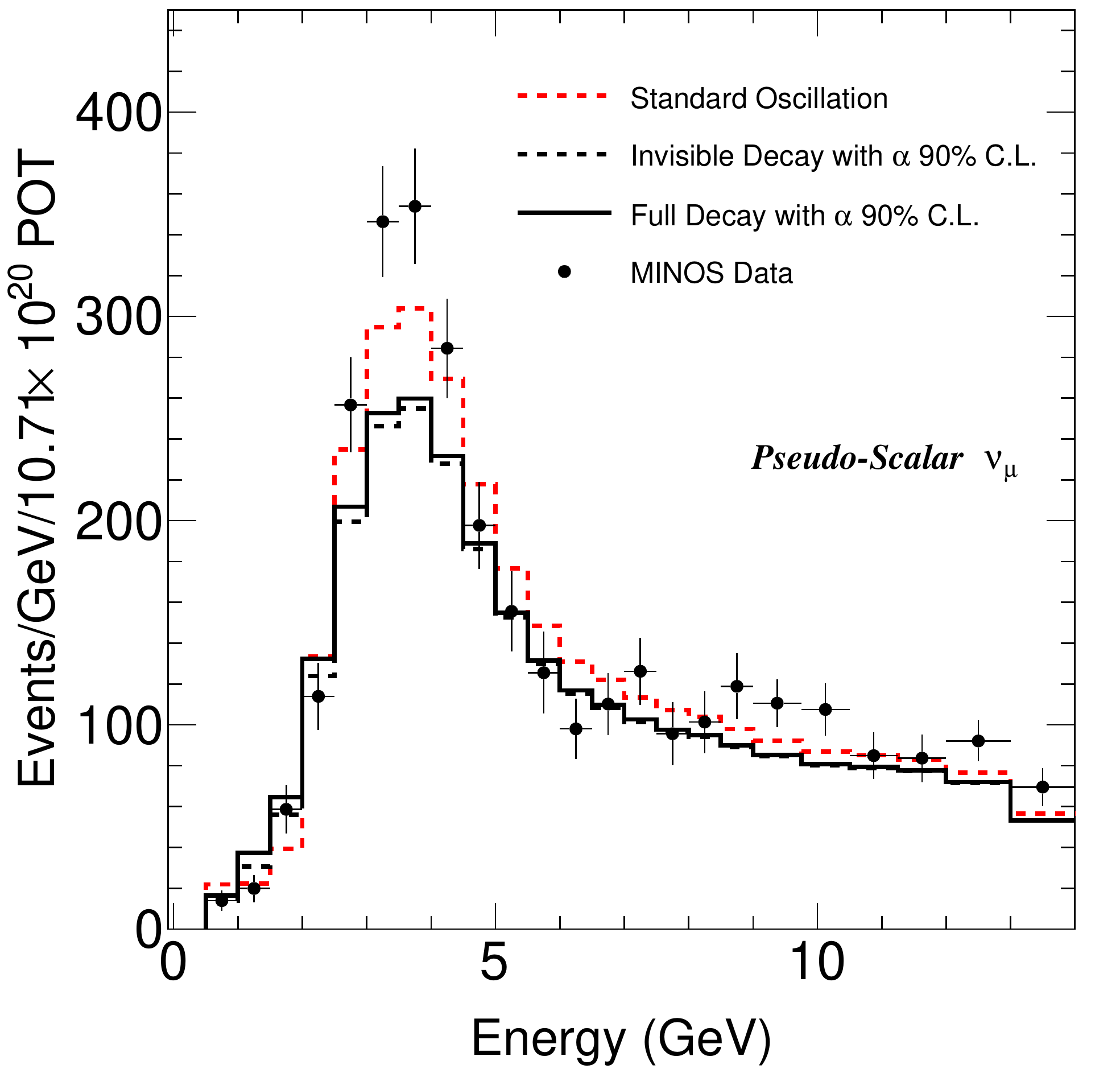} \\
\includegraphics[width=0.45\textwidth]{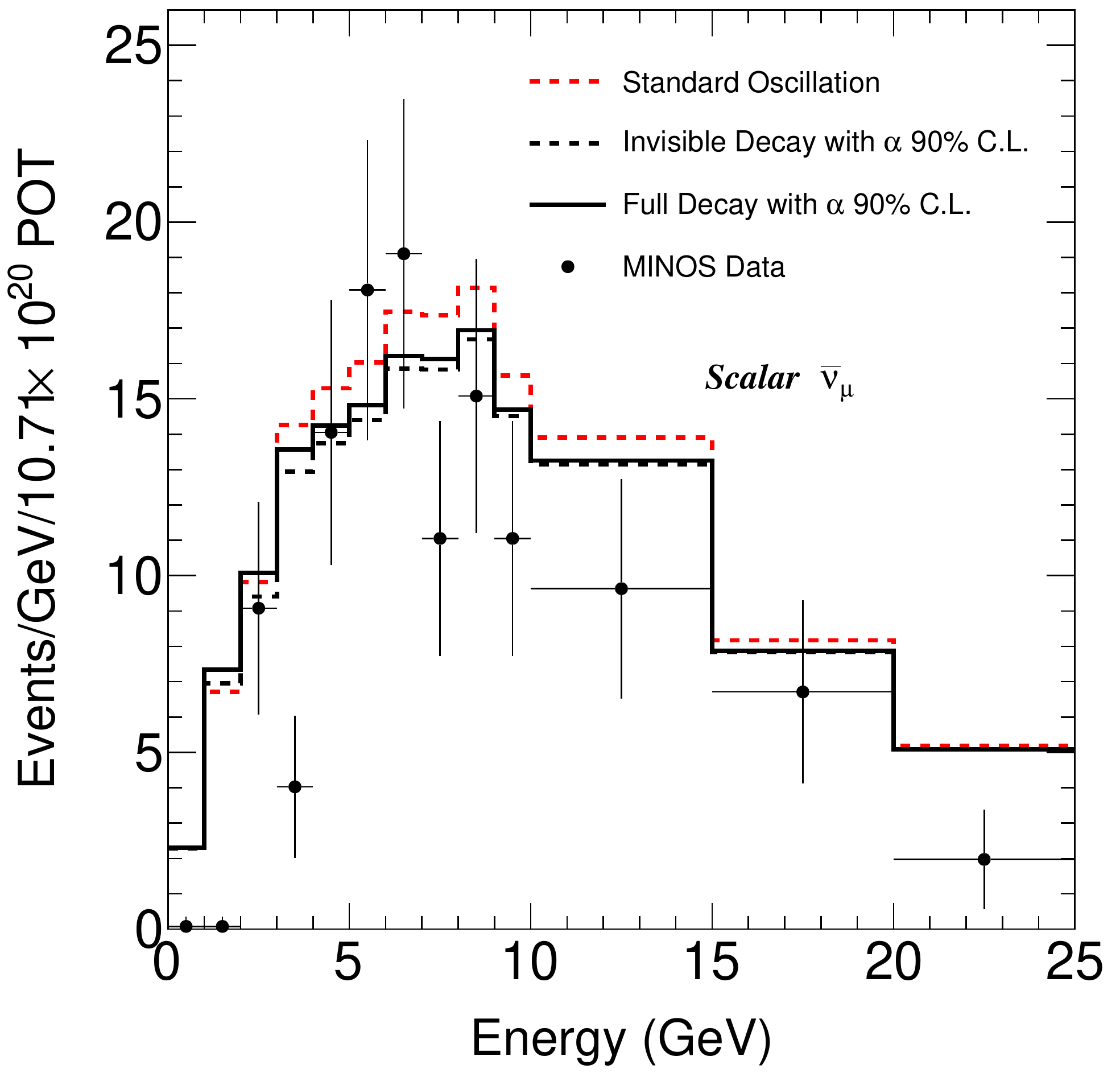} \hfill
\includegraphics[width=0.45\textwidth]{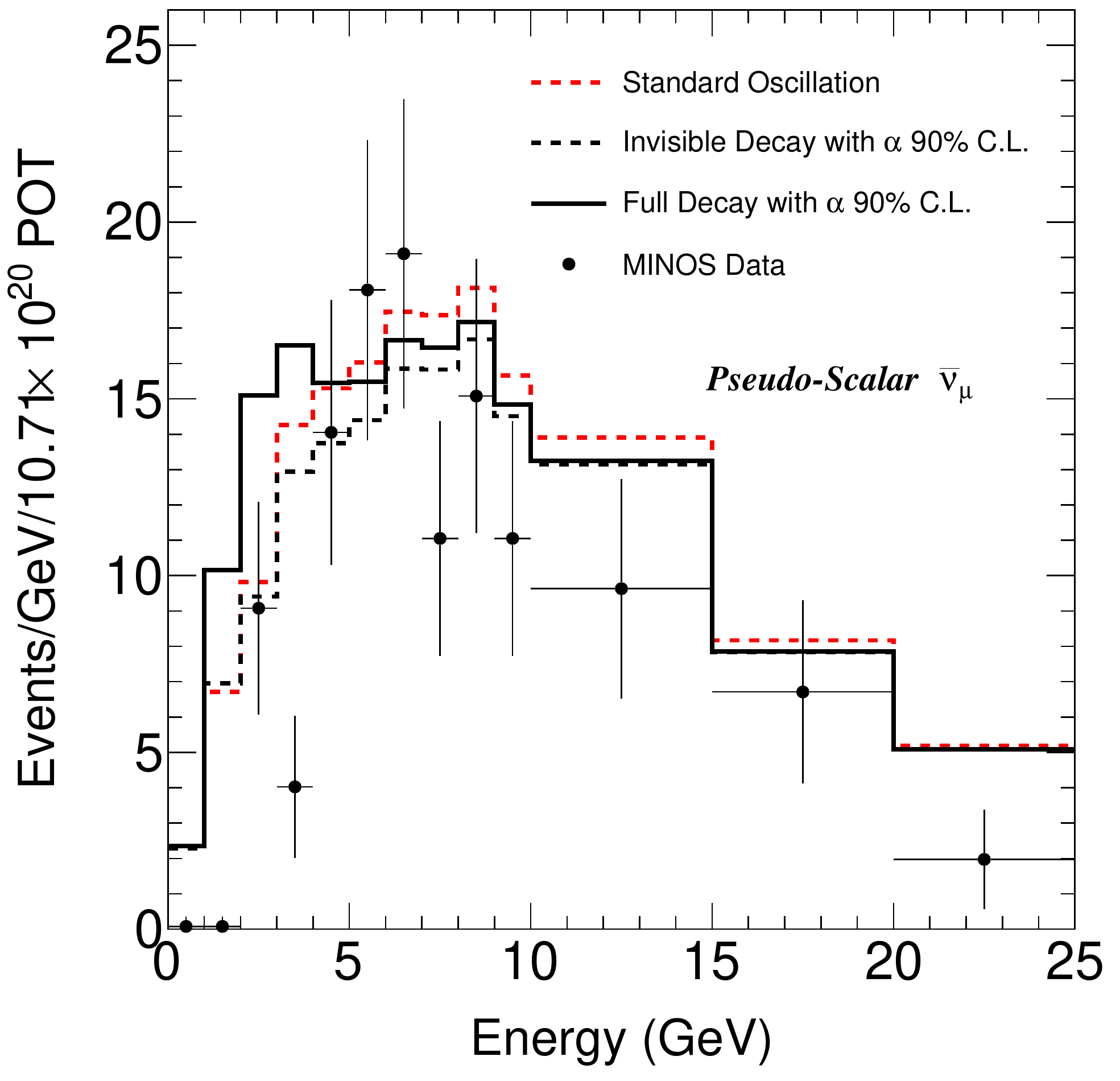}
\caption{
The spectrum of muon and anti-muon neutrinos events is presented for the SO, ID and FD scenarios. We fixed the oscillation parameters in  $\Delta m^2_{32}=2.41\times10^{-3}$~eV$^2$, $s^2_{23}=0.41$, $s^2_{13}=0.0243$, and $\delta_{\rm CP}=0$. We also fixed the decay parameter $\alpha=3.0\times10^{-4}$~eV$^2$ and $m_{\rm light}=0.05$ eV. The MINOS data points were taken from Ref.~\cite{Adamson:2013whj, Cao:2014eca}.
\label{fig:constraints.MINOS}}
\end{figure}
In Figure \ref{fig:constraints.MINOS}, we show the spectrum of muon neutrino (top panels) and antineutrino (bottom panels) disappearance events of FHC mode, for SO, ID and FD scenarios. 
The first (second) column shows the scalar (pseudoscalar) couplings case.
The panels follows the convention presented in Section~\ref{sec:constraints.T2K},
where we have fixed the oscillation parameters 
$s^2_{23}=0.41$, $s^2_{13}=0.0243$, $\Delta m^2_{32}=2.41\times10^{-3}~\text{eV}^2$, which correspond to the best fit of MINOS to standard oscillation model, and $\delta_{\rm CP}=0$, because it is not sensitive in MINOS. When the decay is present, we have set $\alpha=3.0\times10^{-4}~{\rm eV^2}$, and $m_{\text{light}}=0.05~\text{eV}$ to investigate the difference between the scalar and pseudoscalar scenarios in MINOS (see Figure \ref{fig:Ffuncs}).

It is interesting to compare the ID with the FD case to see the effect of adding visible neutrino decay. As explained in the previous Section, and in Figure~\ref{fig:Ffuncs}, when we fixed a large value of $m_{\text{light}}$, this means that for scalar couplings we have a preference 
for $\pm\pm$ transitions. In contrast, pseudoscalar couplings have a clear preference for $\pm\mp$ transitions. This explain why the visible neutrino decay contribution for muon neutrino spectrum is more significant in the 
scalar case. This happens because the NuMI beam configuration was composed of more than $90\%$ of muon neutrinos favoring the transition $\nu_\mu^{(-)}\rightarrow\nu_\mu^{(-)}$. By the other side, looking the muon anti-neutrino spectrum, we see that the contribution of visible neutrino 
decay is higher for pseudoscalar case. Using the same previous reason, this occur because the transition $\nu_\mu^{(-)}\rightarrow\nu_\mu^{(+)}$ is favored for the value of $m_{\text{light}}$ chosen.

The best fit obtained for ID and FD scenario were for an $\alpha$ different from zero. Then, in MINOS, this effect results in an impact on oscillation parameters compared to the one obtained for SO. 
This same effect is observed at Ref.~\cite{Gomes:2014yua}. However, if we take the minimum $\chi^2$ for FD and SO, which are $42.04$ and $45.62$ respectively, and taking into account the data bins and different number of parameters in each scenario, one can demonstrate through the Akaike Information Criterion~\cite{Akaike1, Akaike2} that there is no statistical difference between both models. 

\begin{figure}[!t]
\centering
\includegraphics[width=0.45\textwidth]{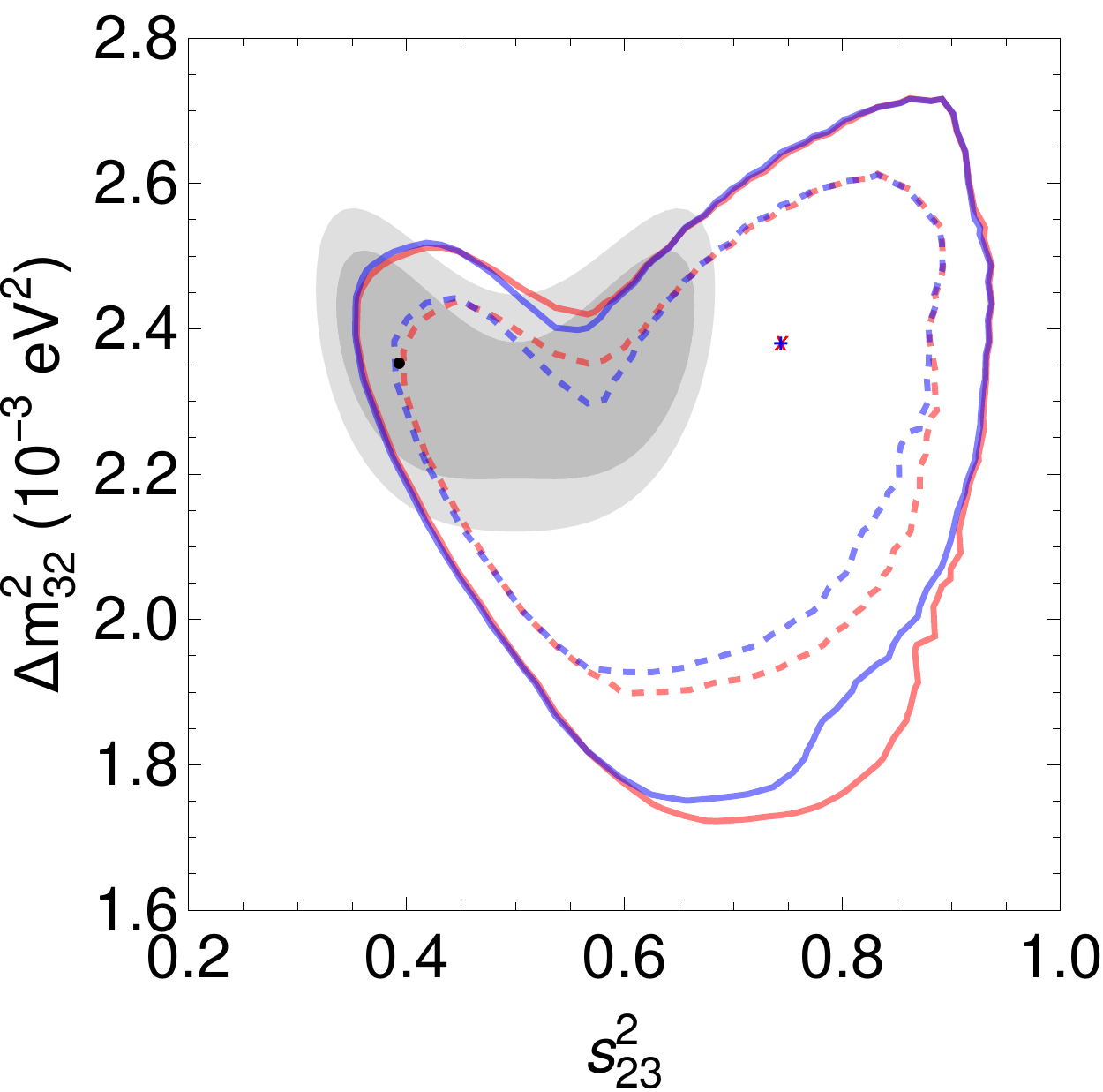} \hfill
\includegraphics[width=0.45\textwidth]{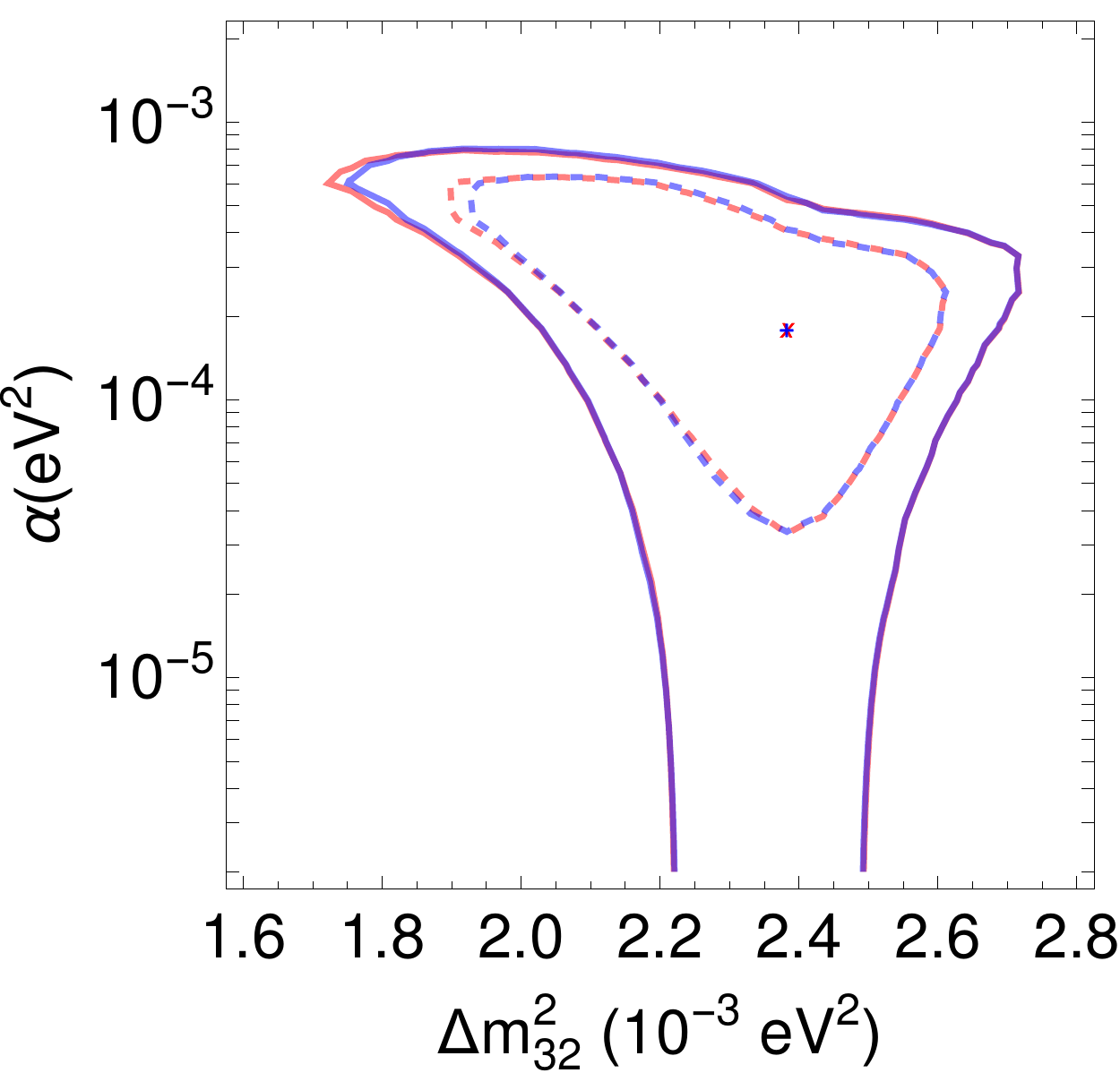}\\
\includegraphics[width=0.45\textwidth]{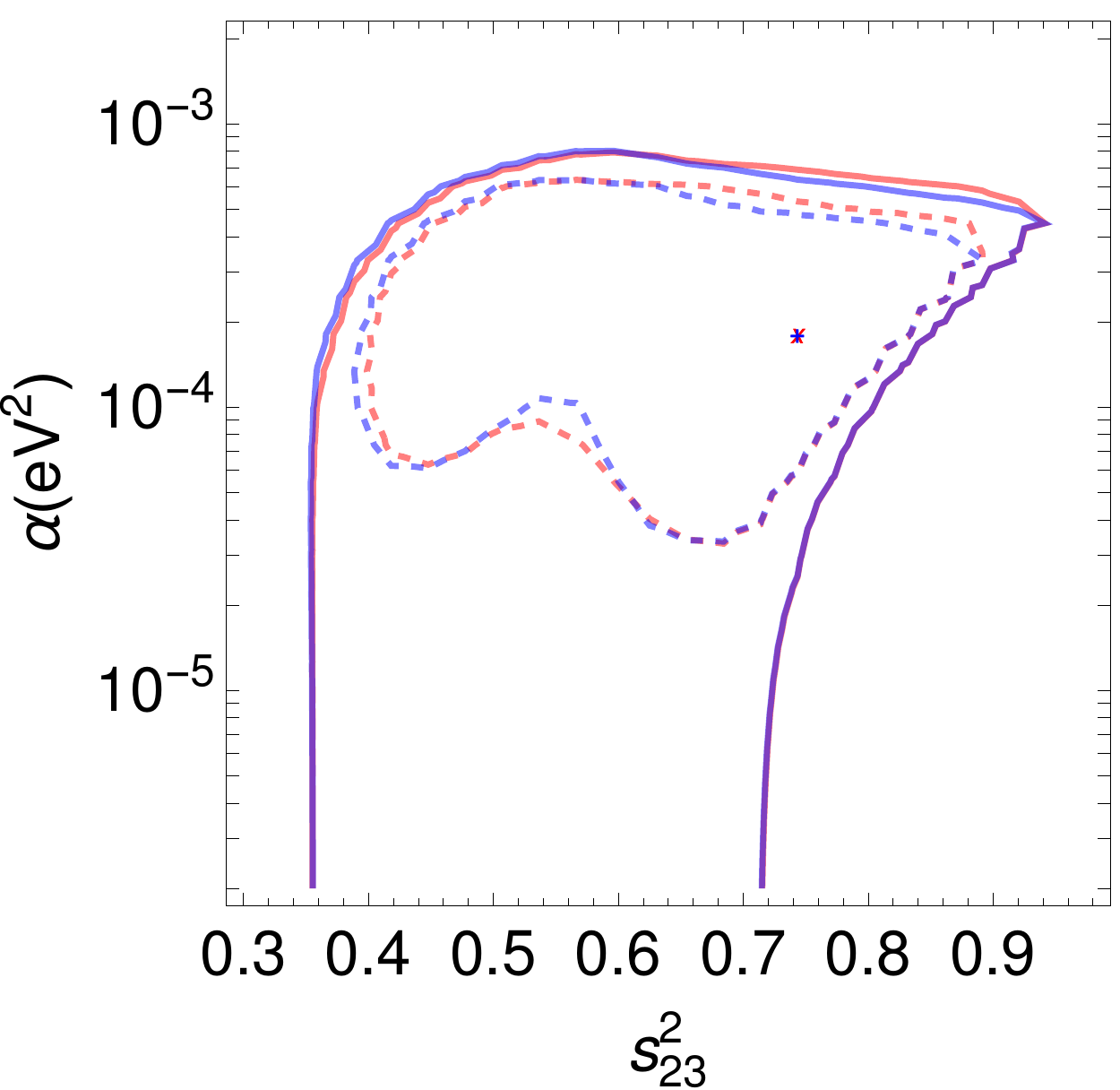} \hfill
\includegraphics[width=0.45\textwidth]{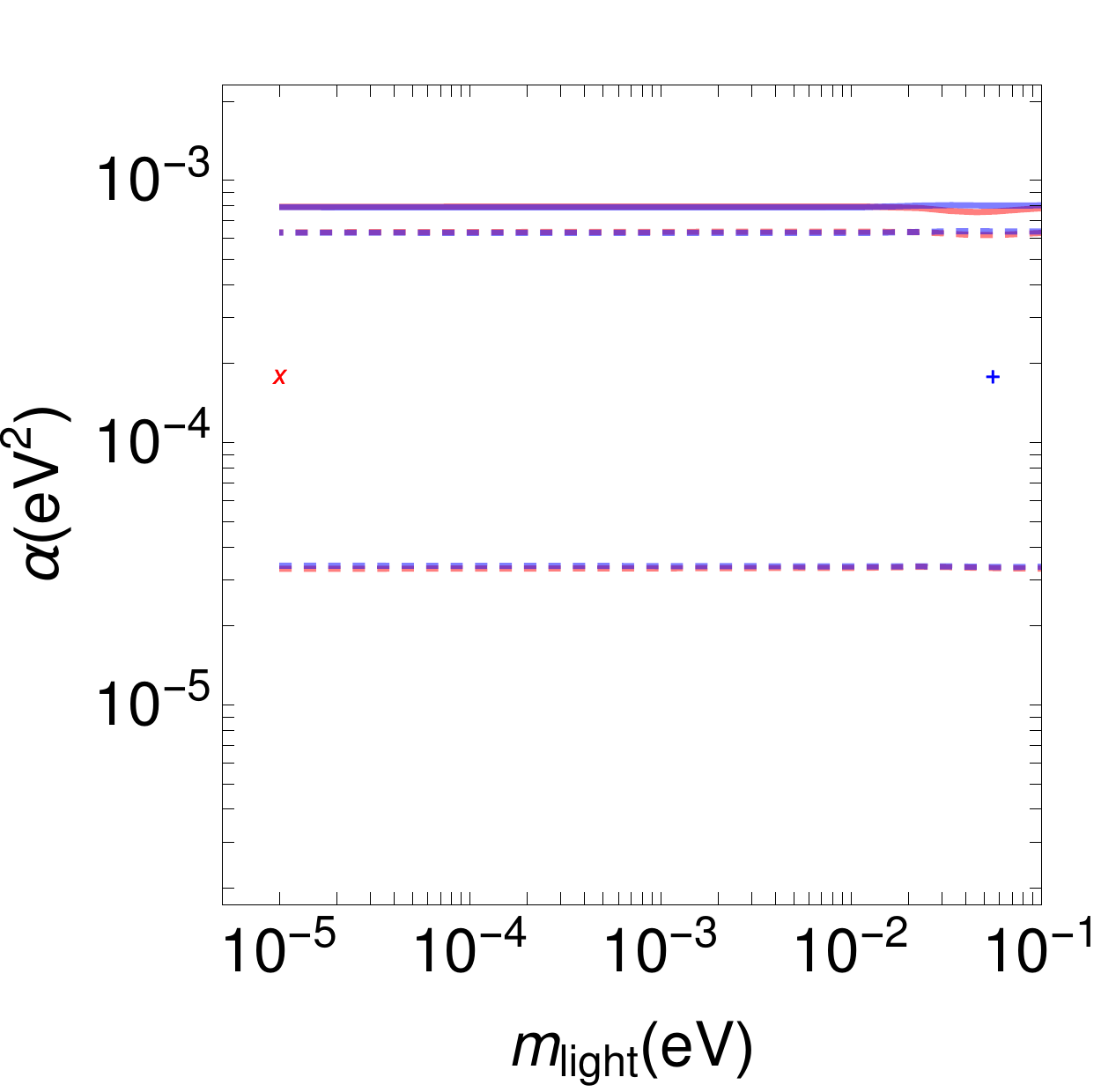}
\caption{\label{fig:allowed.regions.a.MINOS} The allowed regions for the cases scalar and pseudoscalar using the disappearance signals from MINOS at FHC mode. The projections shown are to respect the oscillation parameters and the decay parameter.}
\end{figure}
In Figure \ref{fig:allowed.regions.a.MINOS}, we show the 1$\sigma$ and 90$\%$ C.L. allowed regions for many combination of the relevant parameters, already computing $\nu^{(-)}_{\mu}$
combined with $\nu^{(+)}_{\mu}$. The red (blue) curves refer to scalar (pseudoscalar) couplings. To compare the effect of neutrino decay, we add as well the limit when $\alpha\rightarrow0$, which corresponds to SO scenario (full gray) in Eq.~(\ref{eq:oscdecay}).

The allowed region in the ($s^2_{23}-\Delta m^2_{32}$) plane show that the FD scenario has a significant impact when compared with SO scenario. This means that for MINOS, the contribution of events
coming from visible decay is not enough to constrain the asymmetry at the probability of ID scenario. Another important observation is that we do not find significant differences between scalar and pseudoscalar couplings except by a region in ($s^2_{23}-\Delta m^2_{32}$) plane. We can see it for small values of $\Delta m^2_{32}$ and higher for $s^2_{23}$.

We also present the two-dimensional projections ($\alpha-\Delta m^2_{32}$) and ($\alpha-s^2_{23}$) in Figure~\ref{fig:allowed.regions.a.MINOS}. We can observe that for large values of $\alpha$, the contours increase the range in $\Delta m^2_{32}$, and either allow higher values of $s^2_{23}$.

The right bottom panel in Figure~\ref{fig:allowed.regions.a.MINOS} show the correlation between $\alpha$ and $m_{\text{light}}$. We can conclude by it, that MINOS is not able to distinguish the $\alpha$ constraint for scalar and pseudoscalar couplings.

\subsection{Combination of T2K and MINOS}

\begin{figure}[!t]
\centering
\includegraphics[width=0.45\textwidth]{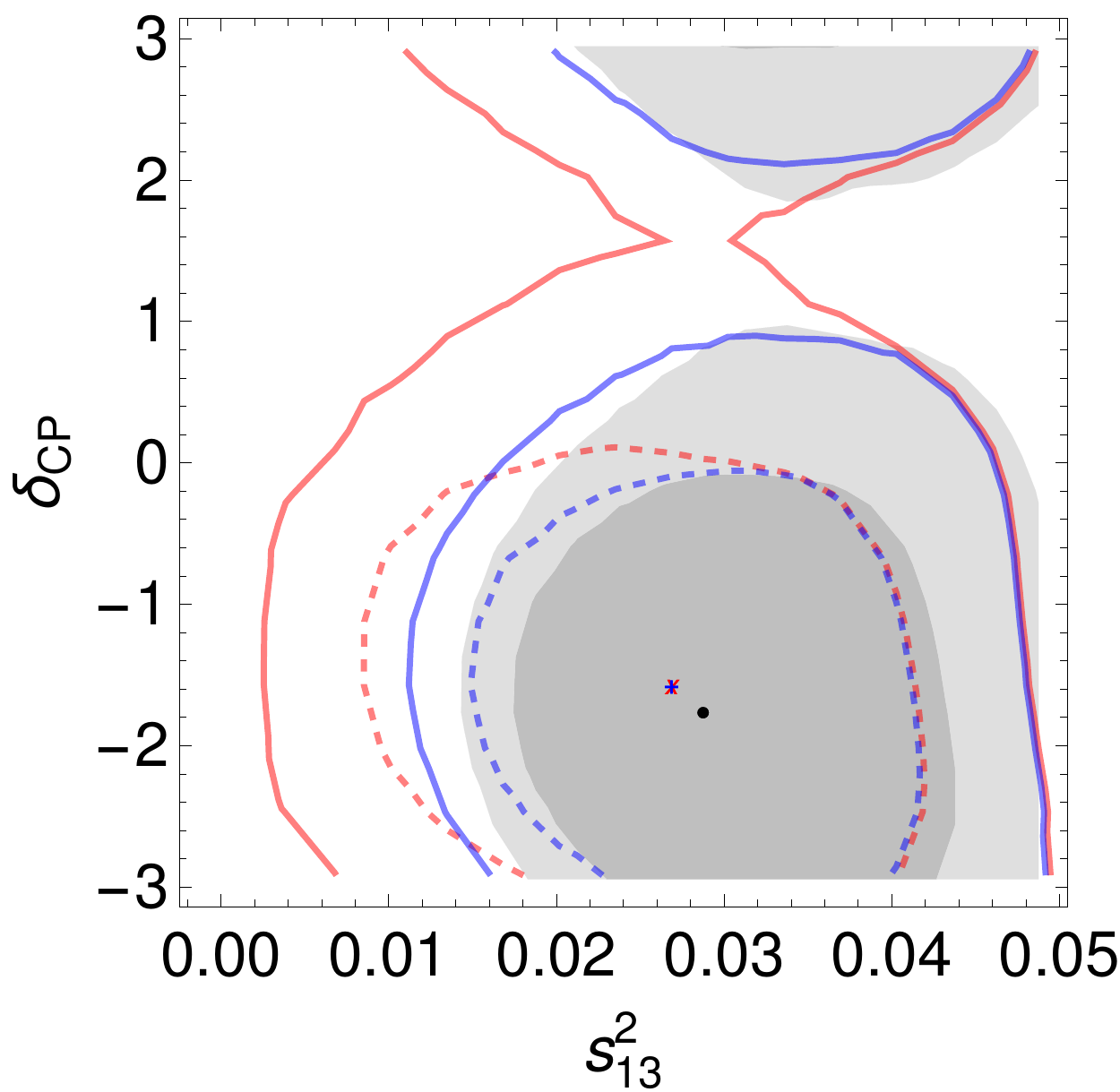} \hfill
\includegraphics[width=0.45\textwidth]{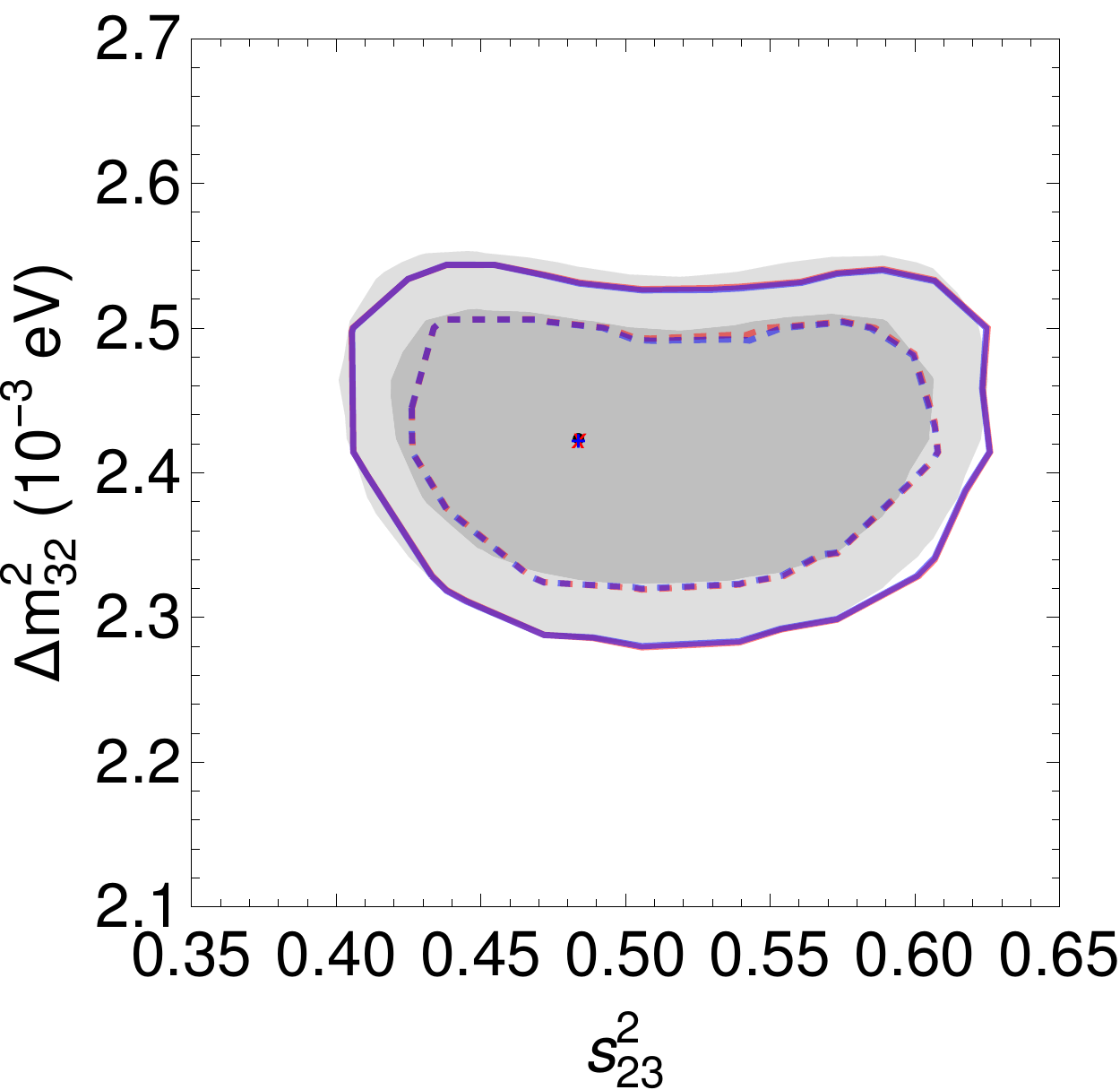} \\
\includegraphics[width=0.45\textwidth]{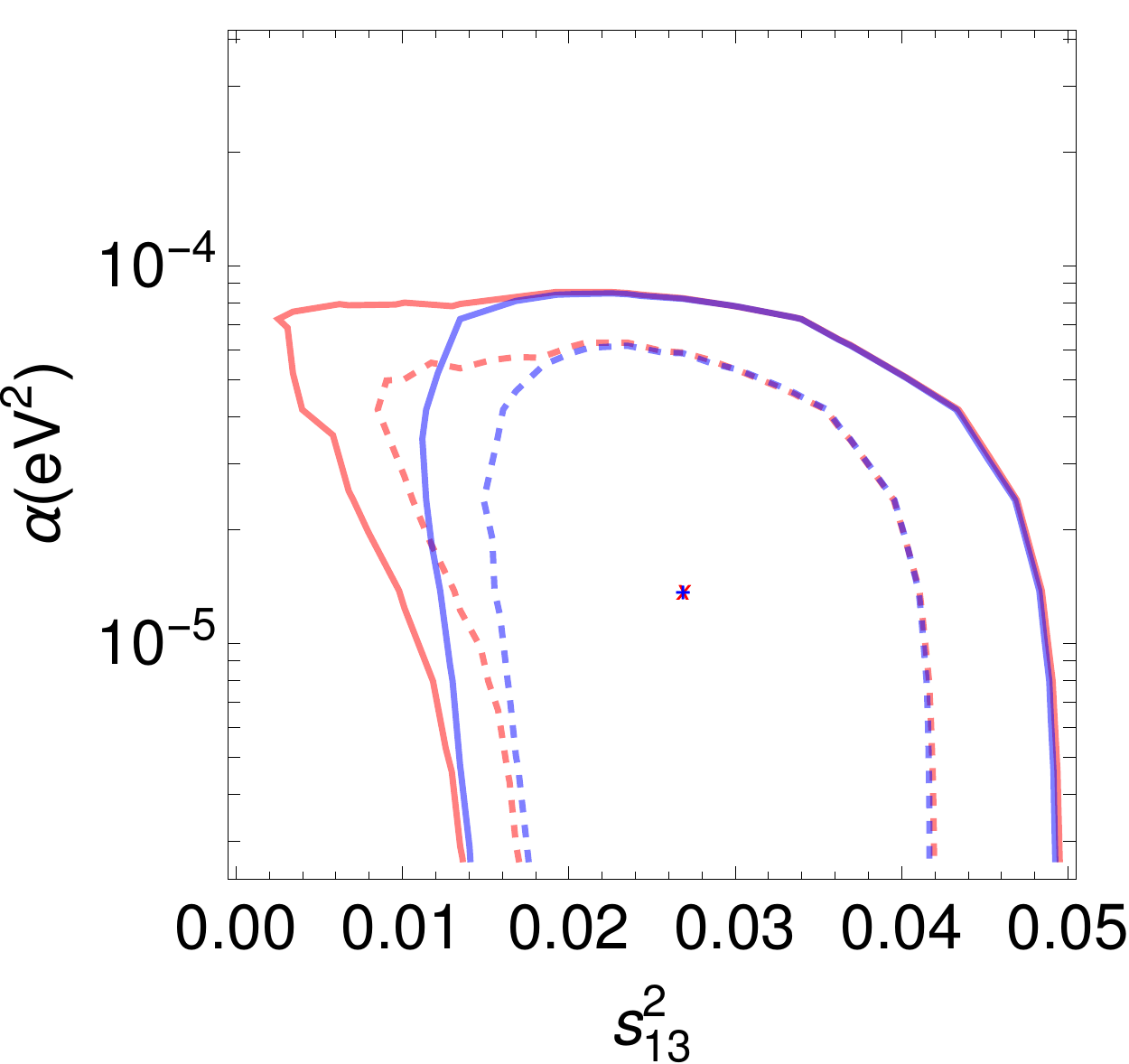} \hfill
\includegraphics[width=0.45\textwidth]{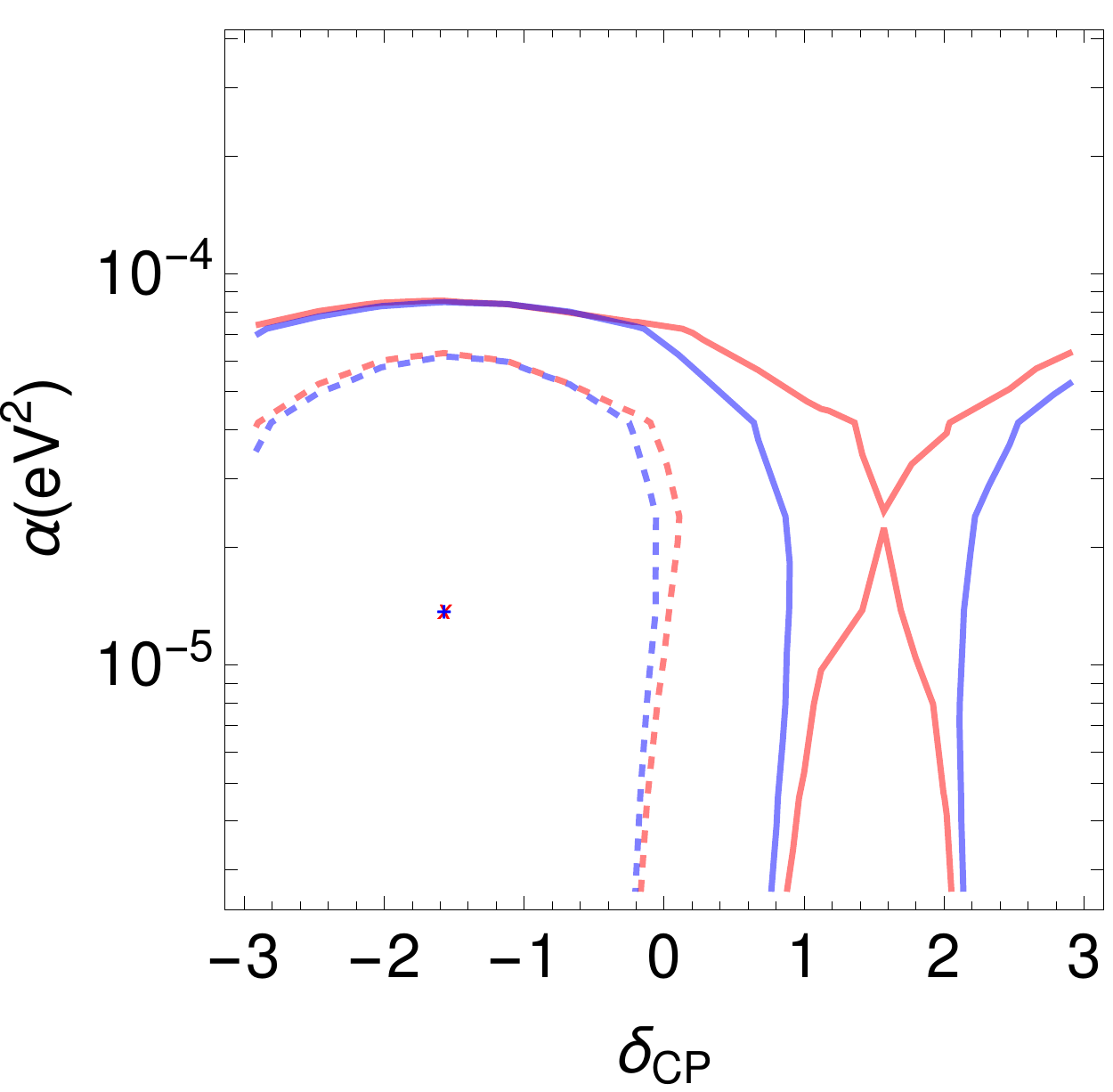}
\caption{As in Figures~\ref{fig:constraints1.nunubar} and \ref{fig:allowed.regions.a.MINOS}, but combining T2K and MINOS data.}
\label{fig:constraints1.T2K+MINOS}
\end{figure}

In order to combine T2K and MINOS data, for every points in the parameter space we have added the respective $\chi^2$ values for each experiment. We show the result in Figure~\ref{fig:constraints1.T2K+MINOS}. The panels follow the same conventions as in Figures~\ref{fig:constraints1.nunubar} and~\ref{fig:allowed.regions.a.MINOS}.

On the top right panel we show the allowed $s^2_{23}-\Delta m^2_{32}$ subspace. We find that the resulting allowed area is not as large as the one in the MINOS-only analysis, in fact, for both scalar and pseudoscalar scenarios, the contours match those for the SO result. This is due to the value of $\alpha$ being strongly constrained by $\nu_e$ appearance, down to values where MINOS has no sensitivity. This is analogous to the T2K-only case, where we saw that no effects were visible for $\nu_\mu$ disappearance. Therefore, the T2K analysis dominates the fit.

On the top left panel, we show the $s^2_{13} - \delta_{\rm CP}$ region. The allowed area in the pseudoscalar scenario is very similar to the T2K-only result. For the scalar scenario, however, when compared to the T2K-only analysis, we find that the regions allow even smaller values of $s^2_{13}$. In addition, in this scenario no value of $\delta_{\rm CP}$ is ruled out at $90\%$~C.L.

The qualitative behaviour observed on the bottom left panel, showing the $s^2_{13}-\alpha$ plane, is very similar to the one of Figure~\ref{fig:constraints1.nunubar}. An evident difference is that, for both couplings, the best-fit for $\alpha$ is increased to a value of $\ord{10^{-5}}$. This implies that the MINOS data shall still relevant when bounding this parameter. Nevertheless, we shall find these best-fit values not to be statistically relevant.

Finally, on the bottom right panel, we can understand why all values of $\delta_{\rm CP}$ are allowed in the scalar coupling scenario. In the SO scenario, if we fix $s_{13}^2$ at the reactor best-fit value, positive $\delta_{\rm CP}$ implies too few (too many) events in the T2K neutrino (antineutrino) run. If we only consider data from the T2K neutrino run, we would find in both FD scenarios that positive $\delta_{\rm CP}$ is incompatible with a vanishing value of $\alpha$. This means that the lesser number of events in the neutrino run, coming from the oscillation contribution, is compensated by the additional contribution from visible decay. This was previously explained in Section~\ref{sec:constraints.T2K} (see the lower panels of Figure~\ref{fig:spectrum.nu}). Nevertheless, the data from the T2K antineutrino run is incompatible with this solution (lower panels of Figure~\ref{fig:spectrum.anu}), such that positive $\delta_{\rm CP}$ is not allowed. 
This is precisely what happened in Figure~\ref{fig:constraints1.nunubar}. However, when adding the MINOS data, we find that the latter pulls the fit towards larger $\alpha$, for all values of $\delta_{\rm CP}$ for both couplings, reintroducing the ambiguity in $\delta_{\rm CP}$ for the scalar case. This ambiguity with $\delta_{\rm CP}$ does not happen for the pseudoscalar case, as we have seen, the incompatibility of  a value of $\alpha$ of $\ord{10^{-5}}$, for $\delta_{\rm CP}=+\pi/2$, in front of antineutrino data is much more serious. One would expect that, if further T2K antineutrino data exhibits the same preference for SO, the positive $\delta_{\rm CP}$ solution would eventually also be excluded for scalar coupling.

\begin{figure}[tbp]
\centering
\includegraphics[width=0.45\textwidth]{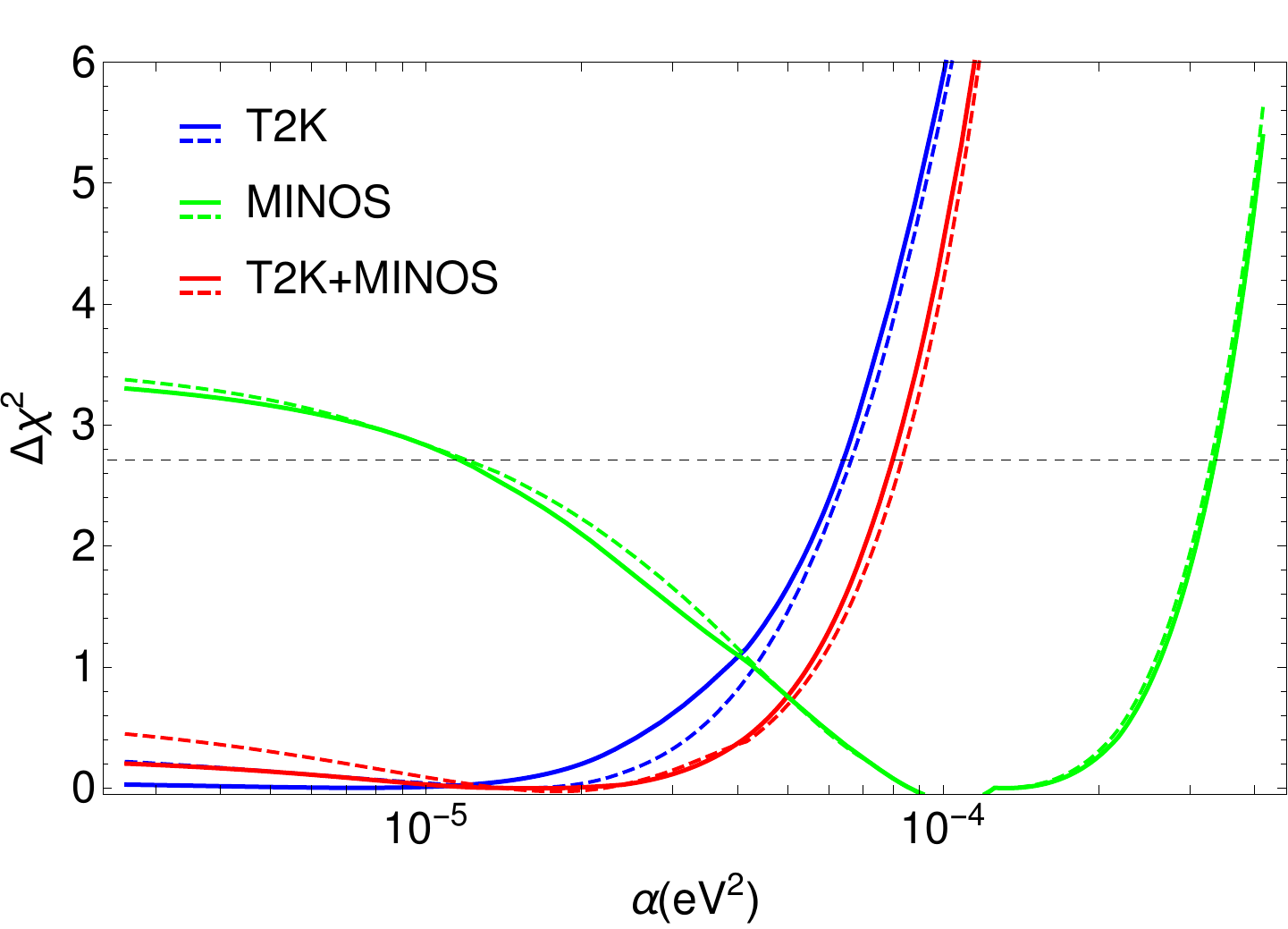} \hfill
\includegraphics[width=0.45\textwidth]{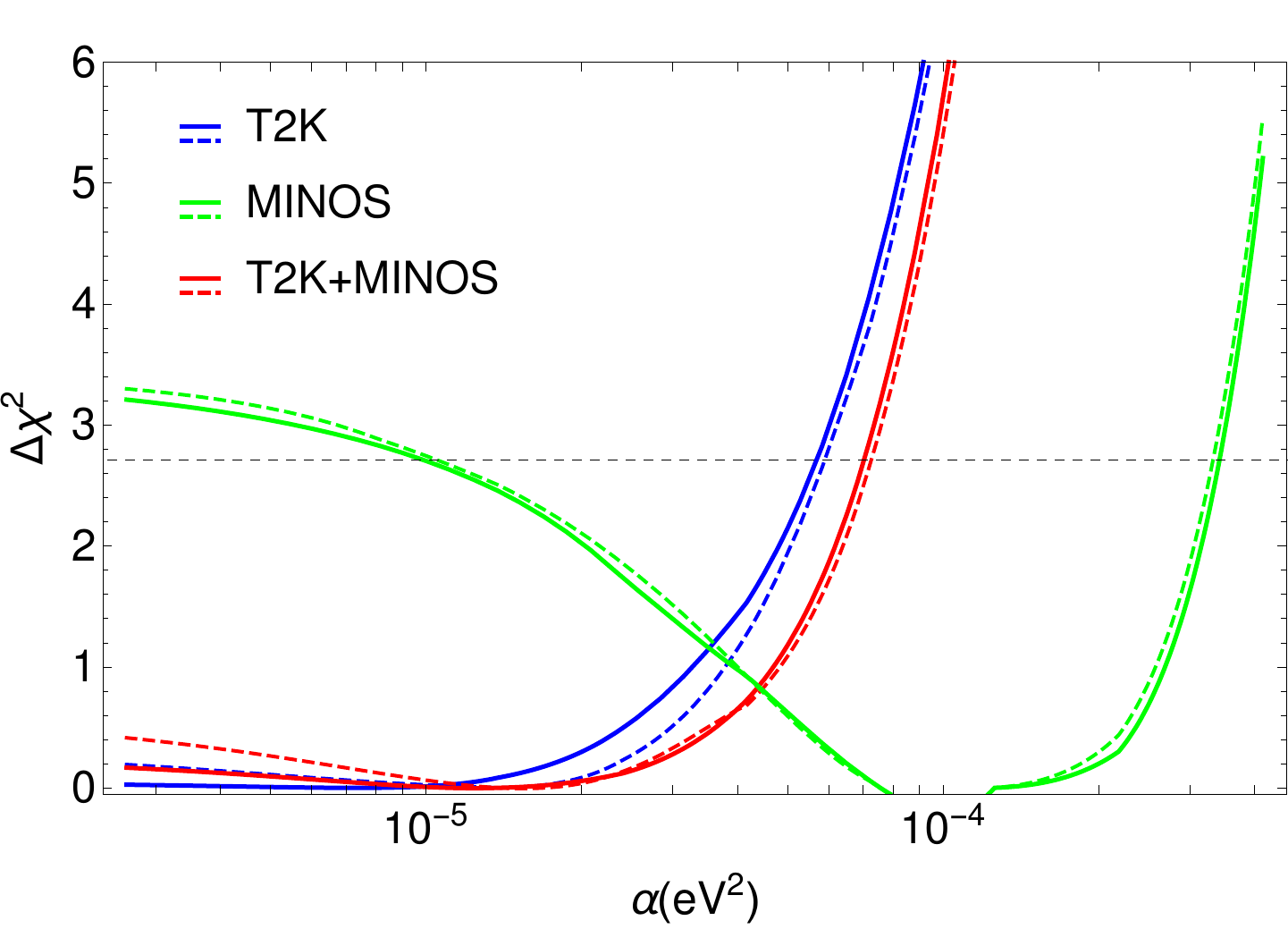}
\caption{Curves showing $\Delta\chi^2$ as a function of $\alpha$ for T2K (blue), MINOS (green) and the full T2K and MINOS combination (red). Dashed lines show the same information, with the additional pull function from reactor data. We show the scalar (pseudoscalar) scenario on the left (right) panel.}
\label{fig:alphachi}
\end{figure}
In Figure~\ref{fig:alphachi} we present show $\Delta\chi^2$ as a function of $\alpha$, for the two scenarios (scalar and pseudoscalar couplings) and three fits (T2K, MINOS and T2K+MINOS) considered. We also add additional curves showing the impact of including reactor data, as explained in Section~\ref{sec:statistics}.

As we can see, the MINOS solution for non-zero $\alpha$ is excluded by T2K data, which  is compatible with a null value for this parameter. Therefore, we place bounds on $\alpha$. The T2K-only $90\%$~C.L. bounds are $\alpha\lesssim5.6\times10^{-5}$~eV$^2$ ($\alpha\lesssim6.3\times10^{-5}$~eV$^2$) for the pseudoscalar (scalar) scenario. As expected, the constraint on this parameter for the T2K+MINOS combination is slightly weaker: $6.9\times10^{-5}$~eV$^2$ ($7.8\times10^{-5}$~eV$^2$) for the pseudoscalar (scalar) case.

The addition of reactor data adds a strong pull factor towards non-zero $s^2_{13}$. However, as the solutions for both couplings involve values of $s^2_{13}$ compatible with reactor data, the addition of the latter does not modify significantly the results.

\section{Conclusions}
\label{sec:conclusions}

In this work we have studied the implications of visible neutrino decay on T2K and MINOS experiments. We have considered that the decay proceeds through the coupling of two neutrinos and a Majoron, which can be either of scalar or pseudoscalar nature. We present our results for both couplings, separately, and describe their effects in detail.

In our analysis we consider neutrino disappearance and appearance channels, both in T2K and MINOS. The results for MINOS point towards a non-zero $\alpha$ as a best-fit solution. Nevertheless, the inclusion of T2K neutrino and antineutrino data excludes this result, being consistent with a vanishing $\alpha$. Consequently, we put bounds on this parameter.

The power of the T2K data lies on the strong influence that visible neutrino decay has on appearance channels. The additional events from the full decay framework are significant even for values of $\alpha$ much smaller than the characteristic $E/L$ where the neutrino experiment has its largest sensitivity. In contrast, for the neutrino disappearance spectrum, the inclusion of visible and invisible neutrino decay components does not differ from what is obtained when considering only the invisible component.

Our results depend on the type of non-zero coupling. For scalar coupling, we have briefly seen that adding reactor data improves the bounds. If we denote the allowed values of $\alpha$, given the coupling $c$, due to the experiment $exp$, by $\alpha^{(c)}_{exp}$, the best constraint we find at $90\%$~C.L.~is:
\begin{align}
 \alpha^{(s)}_{\rm T2K} < 6.3\times10^{-5}{~\rm eV}^2~, & & \alpha^{(p)}_{\rm T2K} < 5.6\times10^{-5}{~\rm eV}^2~, \\
 \alpha^{(s)}_{\rm TK2+MINOS} < 7.8\times10^{-5}{~\rm eV}^2~, & & \alpha^{(p)}_{\rm TK2+MINOS} < 6.9\times10^{-5}{~\rm eV}^2~.
\end{align}

Notice that these bounds depend crucially on the T2K antineutrino data, which is currently in its first years of data taking. We expect that the bounds shall improve with further data.

\acknowledgments

A.M.G.~and J.J.P.~acknowledge funding by the {\it Direcci\'on de Gesti\'on de la Investigaci\'on} at PUCP, through grant DGI-2015-3-0026. They would also like to thank S.~Sanchez for discussions on the use of {\tt GENIE}, as well as O.~A.~D\'iaz in the {\it Direcci\'on de Tecnolog\'ias de Informaci\'on} at PUCP, for implementing our code within the {\tt LEGION} system. A.M.G~is thankful for the Unicamp hospitality and for the support of the FAPESP funding grant 2015/20129-3. A.M.G~and O.L.G.P~are also thankful to the {\it Oficina de Internacionalizaci\'on de la Investigaci\'on} at PUCP.
A.L.G.G. is supported by CAPES, and R.A.G. would like to thanks CNPq grants 461147/2014-9 and 311980/2013-9. O.L.G.P.~is thankful for the PUCP hospitality and partial support under the {\it Programa Latino Americano de F\'isica} (PLAF) from Brazilian Physics Society. He is also thankful for the support of FAPESP funding grant 2014/19164-6 and CNPq research fellowship ~307269/2013-2 and  304715/2016-6 and partial funding from ICTP.

\appendix

  \section{Full Formula for Visible Neutrino Decay}
\label{app:formulae.kin}
  
The $G_{\alpha\beta}^{rs}(E_\alpha,\,E_\beta)$ function describing visible neutrino decay can be derived from the original formula of~\cite{PalomaresRuiz:2005vf}, using the methods of~\cite{Lindner:2001fx}. Assuming the final state neutrinos do not oscillate, the result is:
 \begin{eqnarray}
 \label{eq:appearance}
 G_{\alpha\beta}^{rs}(E_\alpha,\,E_\beta)&=& \sum_{i=2}^{3}\sum_{j=1}^{i-1}\sum_{m=2}^{3}\sum_{n=1}^{m-1}
  \frac{2E_\alpha}{4\alpha_{\langle im\rangle}^2+(\Delta m^2_{im})^2}\sqrt{\frac{d}{dE_\beta}\Gamma(\nu_i^r\to\nu_j^sJ)\frac{d}{dE_\beta}\Gamma(\nu_m^r\to\nu_n^sJ)} \nonumber \\
  &&\bigg{\{}\Re e\left[K_{jinm}^{\beta\alpha sr}\right]\bigg{(}2\alpha_{\langle im\rangle}\left[\cos\left(\frac{\omega^{sr}_{ijmn}}{2} \right)-\cos\Delta^{sr+}_{ijmn}\exp\left(-\frac{\alpha_{\langle im\rangle}L}{E_\alpha}\right)\right] \nonumber \\
  && \left.\qquad\qquad\qquad\quad-\Delta m^2_{im}\left[\sin\left(\frac{\omega^{sr}_{ijmn}}{2} \right)+\sin\Delta^{sr-}_{ijmn}\exp\left(-\frac{\alpha_{\langle im\rangle}L}{E_\alpha}\right)\right]\right) \nonumber \\
  &&-\Im m\left[K_{jinm}^{\beta\alpha sr}\right]\bigg{(}2\alpha_{\langle im\rangle}\left[\sin\left(\frac{\omega^{sr}_{ijmn}}{2} \right)-\sin\Delta^{sr+}_{ijmn}\exp\left(-\frac{\alpha_{\langle im\rangle}L}{E_\alpha}\right)\right] \nonumber \\
  && \left.\qquad\qquad\qquad\quad+\Delta m^2_{im}\left[\cos\left(\frac{\omega^{sr}_{ijmn}}{2} \right)-\cos\Delta^{sr-}_{ijmn}\exp\left(-\frac{\alpha_{\langle im\rangle}L}{E_\alpha}\right)\right]\right)\bigg{\}} \nonumber\\
  \end{eqnarray}
 Here, we have $K_{jinm}^{\beta\alpha sr}=U^{(s)}_{\beta j}U^{(r)*}_{\alpha i}U^{(s)*}_{\beta n}U^{(r)}_{\alpha m}$.  
Notice that, following Eq.~(\ref{eq:PMNS}), the PMNS matrix has been defined without the Majorana phases. In Eq.~(\ref{eq:appearance}), the latter are taken into account within the parameter $\omega_{ijmn}^{sr}=\gamma^s_{jj}-\gamma^{r}_{ii}-\gamma^s_{nn}+\gamma^r_{mm}$. We also denote:
 \begin{eqnarray}
 \alpha_{\langle im\rangle} &=& \frac{\alpha_i+\alpha_m}{2}~, \\
 \Delta^{sr\pm}_{ijmn} &=& \frac{\Delta m^2_{im}}{2E_\alpha}L\pm\frac{\omega^{sr}_{ijmn}}{2}~.
\end{eqnarray}

 If we only have one unstable neutrino, this expression can be considerably simplified. We chose $\nu_3^r$ to be unstable, which means we fix $i=m=3$ in Eq.~(\ref{eq:appearance}), so we obtain:
 \begin{multline}
 \label{eq:appearance.one}
 G_{\alpha\beta}^{rs}(E_\alpha,\,E_\beta)= \left(\frac{1-e^{-\alpha_3 L/E_\alpha}}{\alpha_3/E_\alpha}\right)\left|U_{\alpha 3}^{(r)}\right|^2\Bigg{\{}
  \sum_{j=1}^{2}\left|U_{\beta j}^{(s)}\right|^2\frac{d}{dE_\beta}\Gamma(\nu_3^r\to\nu_j^sJ)
 \\  
  + \left.2\Re e\left[U_{\beta2}^{(s)}U_{\beta1}^{(s)*}\exp\left(i\frac{\omega^{sr}_{3231}}{2}\right)\right]\sqrt{\frac{d}{dE_\beta}\Gamma(\nu_3^r\to\nu_1^sJ)\frac{d}{dE_\beta}\Gamma(\nu_3^r\to\nu_2^sJ})\right\}~.
 \end{multline}
 
Furthermore, if only one decay channel is allowed (for example, $\nu_3^r\to\nu_1^sJ$, obtained by fixing $j=n=1$ in Eq.~(\ref{eq:appearance})), we have:
 \begin{equation}
 \label{eq:appearance.3to1}
 G_{\alpha\beta}^{rs}(E_\alpha,\,E_\beta)= \left(\frac{1-e^{-\alpha_3 L/E_\alpha}}{\alpha_3/E_\alpha}\right)\left|U^{(r)}_{\alpha 3}\right|^2 \left|U^{(s)}_{\beta 1}\right|^2\frac{d}{dE_\beta}\Gamma(\nu_3^r\to\nu_1^sJ)~.
 \end{equation}

One must note that the previous formulae have been written in the convention $m_1<m_2<m_3$, which coincides with the normal hierarchy. To obtain results for the inverted hierarchy, one either must write $m_1\to m_3$, $m_2\to m_1$ and $m_3\to m_2$, or modify the full PMNS matrix such that:
\begin{equation}
 U\cdot{\rm diag}\left(1,\,e^{i\tfrac{\gamma_{22}}{2}},\,e^{i\tfrac{\gamma_{33}}{2}}\right)\to U\cdot\left(\begin{array}{ccc} 0 & 1 & 0 \\ 0 & 0 & 1 \\ 1 & 0 & 0 \end{array}\right)\cdot{\rm diag}\left(e^{i\tfrac{\gamma_{33}}{2}},\,1,\,e^{i\tfrac{\gamma_{22}}{2}}\right)~.
\end{equation}

\section{Formulae for Visible Neutrino Decay Rates}
\label{app:formulae.dyn}

In the following, we shall concentrate on the differential decay width $d\Gamma(\nu_i^r\to\nu_f^s J)/dE_\beta$, for the general case where several decay channels exist. Our procedure follows~\cite{Lindner:2001fx}, where we have a neutrino - Majoron coupling:
 \begin{equation}
  \mathcal{L}_{\rm int}=\frac{(g_s)_{ij}}{2}\bar\nu_i\nu_jJ+i\frac{(g_p)_{ij}}{2}\bar\nu_i\gamma_5\nu_j J~,
 \end{equation}
where the neutrinos are on the mass eigenstates basis.
 
  With this, the partial decay rates of a neutrino with energy $E_\alpha$ are~\cite{Lindner:2001fx,Kim:1990km}:
  \begin{subequations}
 \label{eq:Gammai}
 \begin{eqnarray}
   \Gamma(\nu_i^{(\pm)}\to\nu_f^{(\pm)} J)&=&\frac{m_i^2}{16\pi}\frac{1}{x_{if}E_\alpha}\bigg{[}(g_s)^2_{if}\,f(x_{if})
   +(g_p)^2_{if}\,g(x_{if})\bigg{]}~, 
    \\
   \Gamma(\nu_i^{(\pm)}\to\nu^{(\mp)}_fJ)&=&\frac{m_i^2}{16\pi}\frac{1}{x_{if}E_\alpha}\left((g_s)^2_{if}+(g_p)^2_{if}\right)k(x_{if})~,
  \end{eqnarray}
  \end{subequations}
where $x_{if}=m_i/m_f>1$, and:
\begin{subequations}
\begin{eqnarray}
 f(x) &=& \frac{x}{2}+2+\frac{2}{x}\log x-\frac{2}{x^2}-\frac{1}{2x^3}~, \\
 g(x) &=& \frac{x}{2}-2+\frac{2}{x}\log x+\frac{2}{x^2}-\frac{1}{2x^3}~, \\
 k(x) &=& \frac{x}{2}-\frac{2}{x}\log x-\frac{1}{2x^3}~.
\end{eqnarray}
\end{subequations}
 
 For our description of neutrino decay, we need the differential decay rate:
 \begin{equation}
  \label{eq:diffrate}
  \frac{d}{dE_\beta}\Gamma(\nu_i^r\to\nu_f^s J)=\frac{m_i m_f}{4\pi E_\alpha^2}\left(1-\frac{m_i^2}{E_\alpha^2}\right)^{-1/2}\left|\mathcal{M}(\nu_i^r\to\nu_f^s J)\right|^2 \Theta(E_\alpha,\,E_\beta)~.
 \end{equation}
The $\Theta(E_\alpha,\,E_\beta)$ function fixes the angle between the momenta of the incoming and outgoing neutrinos:
\begin{equation}
 \cos\theta=\frac{2E_\alpha E_\beta-(m_i^2+m_f^2)}{2|\vec{p}_i||\vec{p}_f|}~,
\end{equation}
which effectively imposes the following bound on $E_f$:
\begin{equation}
  \frac{E_\alpha}{2}\left(1+\frac{m^2_i}{m^2_f}\right)-\frac{|\vec{p}_i|}{2}\left(1-\frac{m^2_i}{m^2_f}\right)\leq E_\beta \leq
  \frac{E_\alpha}{2}\left(1+\frac{m^2_i}{m^2_f}\right)+\frac{|\vec{p}_i|}{2}\left(1-\frac{m^2_i}{m^2_f}\right)~.
 \end{equation}
 On the relativistic limit, this reduces to:
 \begin{equation}
 \label{eq:efbound}
 \frac{E_\alpha}{x^2_{if}}\leq E_\beta\leq E_\alpha~.
 \end{equation}
 Thus, in terms of the Heaviside function $\Theta_H$, we have:
 \begin{equation}
  \Theta(E_\alpha,\,E_\beta)=\Theta_H(E_\alpha-E_\beta)\,\Theta_H(x_{if}^2E_\beta-E_\alpha)~.
 \end{equation}
 The neutrino decay amplitudes $\mathcal{M}(\nu_i^{(\pm)}\to\nu_f^{(\pm)} J)$ can be computed as a function of  $(g_p)^2_{if} $ and $(g_s)^2_{if}$, the neutrino masses and the neutrino energies~\cite{Lindner:2001fx,Kim:1990km}. 
 These can be recasted using the partial width of neutrinos, given by Eqs.~(\ref{eq:Gammai}). We first define:
 \begin{equation}
  \alpha_{if}=E_\alpha\left(\Gamma(\nu_i\to\nu_fJ)+\Gamma(\nu_i\to\bar\nu_fJ)\right)~,
  \label{eq3}
 \end{equation}
 such that $\alpha_i=E_\alpha\Gamma_i=\sum_f\alpha_{if}$.  We can invert the Eq.~(\ref{eq3}) and solve for one coupling, such that either the scalar or the pseudoscalar couplings is written: 
\begin{subequations}
 \label{eq:couplings}
 \begin{eqnarray}
  (g_p)^2_{if} &=& \frac{16\pi\alpha_{if}}{m_i^2}\frac{x_{if}^4}{(x_{if}+1)(x_{if}-1)^3}-(g_s)^2_{if}\left(\frac{x_{if}+1}{x_{if}-1}\right)^2~, \\
  (g_s)^2_{if} &=& \frac{16\pi\alpha_{if}}{m_i^2}\frac{x_{if}^4}{(x_{if}+1)^3(x_{if}-1)}-(g_p)^2_{if}\left(\frac{x_{if}-1}{x_{if}+1}\right)^2~. 
 \end{eqnarray}
\end{subequations}
Using Eqs.~(\ref{eq:couplings}), we can re-write the expression of the amplitude $\mathcal{M}(\nu_i^{(\pm)}\to\nu_f^{(\pm)} J)$~\cite{Lindner:2001fx,Kim:1990km} in terms of $\alpha_{ij}$ and one coupling. For example, in terms of $(g_{if})$, we have: 
  \begin{eqnarray}
  \left|\mathcal{M}(\nu_i^{(\pm)}\to\nu_f^{(\pm)} J)\right|^2&=&\frac{x_{if}}{(x_{if}-1)^2}\left[\frac{4\pi\alpha_{if}}{m_i^2}\left(\frac{x_{if}^3}{x_{if}^2-1}\right)(A-2)+(g_s)_{if}^2\left(\frac{1}{x_{if}}+x_{if}-A\right)\right] \nonumber \\
  \left|\mathcal{M}(\nu_i^{(\pm)}\to\nu_f^{(\mp)} J)\right|^2&=&\frac{x_{if}}{(x_{if}-1)^2}\left[\frac{4\pi\alpha_{if}}{m_i^2}\left(\frac{x_{if}^3}{x_{if}^2-1}\right)-(g_s)_{if}^2\right]\left(\frac{1}{x_{if}}+x_{if}-A\right)~.
 \end{eqnarray}
Alternatively, in terms of $(g_p)_{if}$, we find:
 \begin{eqnarray}
  \left|\mathcal{M}(\nu_i^{(\pm)}\to\nu_f^{(\pm)} J)\right|^2&=&\frac{x_{if}}{(x_{if}+1)^2}\left[\frac{4\pi\alpha_{if}}{m_i^2}\left(\frac{x_{if}^3}{x_{if}^2-1}\right)(A+2)-(g_p)_{if}^2\left(\frac{1}{x_{if}}+x_{if}-A\right)\right]  \nonumber \\
  \left|\mathcal{M}(\nu_i^{(\pm)}\to\nu_f^{(\mp)} J)\right|^2&=&\frac{x_{if}}{(x_{if}+1)^2}\left[\frac{4\pi\alpha_{if}}{m_i^2}\left(\frac{x_{if}^3}{x_{if}^2-1}\right)+(g_p)_{if}^2\right]\left(\frac{1}{x_{if}}+x_{if}-A\right)~.
 \end{eqnarray}
For both cases, we have defined:
\begin{equation}
 A=\frac{1}{x_{if}}\frac{E_i}{E_f}+x_{if}\frac{E_f}{E_i}~.
\end{equation}

These results are very useful, as they allow us to simplify the differential decay width. If we go to the limit where one coupling is zero, the other coupling is automatically determined by the value of $\alpha_{if}$. If, in addition, there exists only one allowed decay channel, then the whole process is governed by $\alpha_i$.

These sort of considerations lead us to the principal formula used in this work, Eq.~(\ref{eq:simpleG}).

\section{Formulae for T2K and MINOS}
\label{app:expformulae}
 
In order to obtain the number of events within an energy bin, one needs to multiply Eq.~(\ref{eq:basic}) by the $\nu_{\beta}^s$ cross-section and perform the $E_\beta$ integral considering the particular bin width, efficiency and energy resolution function. We calculate the number of events in the energy bin $i$, with chirality $s$ and going through interaction $int$, using:
\begin{equation}
\label{eq:events}
N^{s,\rm int}_{i,\beta}=
 \int dE_\beta\,K^{\rm int}_i(E_\beta)\,\sigma^{s,\rm int}_\beta(E_\beta)\frac{dN^s_\beta}{dE_\beta}~,
\end{equation}
where $\sigma_\beta^{s,\rm int}(E_\beta)$ denotes the appropriate cross section, and the ``detection kernel'' $K^{\rm int}_i(E_\beta)$ for bin $i$ is defined:
\begin{equation}
 K^{\rm int}_i(E_\beta)=\int_{E_{\rm i, min}}^{E_{\rm i, max}}dE_{\rm bin}\,\epsilon^{\rm int}_\beta(E_{\rm bin})\,
 R(E^{\rm int}_{\beta,\rm rec}-E_\beta,\varsigma_\beta^{\rm int})~.
\end{equation}
Here, $\epsilon^{\rm int}_\beta(E_{\rm bin})$ denotes the detector efficiency, while the resolution function $R(E^{\rm int}_{\beta,\rm rec}-E_\beta,\varsigma_\beta^{\rm int})$ reflects our capacity to reconstruct the true neutrino energy $E_\beta$. In both experiments we model this using a gaussian function:
\begin{equation}
 R(\Delta E,\varsigma_\beta^{\rm int})=\frac{1}{\sqrt{2\pi}\varsigma_\beta^{\rm int}}\exp\left[-\frac{(\Delta E)^2}{2(\varsigma_\beta^{\rm int})^2}\right]~.
\end{equation}
The variable $E^{\rm int}_{\beta,\rm rec}$ is related to the bin energy $E_{\rm bin}$ by a possible energy shift, and $\varsigma^{\rm int}_\beta$ denotes the energy resolution width.

\subsection{T2K}
\label{app:t2kformulae}

For T2K, the neutrino fluxes appearing in Eq.~(\ref{eq:basic}) are obtained from~\cite{Abe:2015awa}, and the cross sections are taken from {\tt GENIE}~\cite{Andreopoulos:2009rq}.

For $\nu_\mu$ disappearance, we take the shift and resolution width $\varsigma_\mu^{\rm int}$ from~\cite{Machado:2011ar}:
\begin{align}
 E^{\rm CCQE}_{\mu,\rm rec} &= E_{\rm bin}~, & E^{\rm CCnQE}_{\mu,\rm rec} &= E_{\rm bin}-0.34\,{\rm GeV}~, \\
 \varsigma_\mu^{\rm CCQE}&=0.085~{\rm GeV}~, & \varsigma_\mu^{\rm CCnQE}&=0.130~{\rm GeV}~.
\end{align}
with equal values for neutrino and antineutrino channels.

For $\nu_e$ appearance, we use the ``migration matrix'' from~\cite{Machado:2013kya}, which is equivalent to using:
\begin{align}
 E^{\rm CCQE}_{e,\rm rec} &= E_{\rm bin}-(0.025 E_{\beta}-3.75\times10^{-3}{\,\rm GeV}) &
 E^{\rm CCnQE}_{e,\rm rec} &= E_{\rm bin}-(0.325E_{\beta}+0.146\,{\,\rm GeV}) \\
 \varsigma_e^{\rm CCQE}&=0.065E_\beta+0.049~{\rm GeV}~, & \varsigma_e^{\rm CCnQE}&=0.2E_\beta-0.04~{\rm GeV}~, \\
 E^{\rm CCQE}_{\bar e,\rm rec} &= E_{\rm bin}-0.02{\,\rm GeV}~, &
 E^{\rm CCnQE}_{\bar e,\rm rec} &= E_{\rm bin}-(0.2E_\beta+0.16\,{\,\rm GeV})~, \\
 \varsigma_{\bar e}^{\rm CCQE}&=0.015E_\beta+0.049~{\rm GeV}~, & \varsigma_{\bar e}^{\rm CCnQE}&=0.1E_\beta+0.045~{\rm GeV}~.
\end{align}

For NC events, we fit the shape of the spectra shown in~\cite{ICHEP-T2K-2016,Salzgeber:2015gua}. We set:
\begin{align}
 E^{\rm NC}_{\mu,\rm rec} &= E_{\rm bin}-0.310{\,\rm GeV}~, & E^{\rm NC}_{e,\rm rec} &= E_{\rm bin}-0.265{\,\rm GeV}~ \\
 \varsigma^{\rm NC}_\mu &= 0.06{\,\rm GeV}~, & \varsigma^{\rm NC}_e &= 0.08{\,\rm GeV}~.
\end{align}

Finally, the detector efficiency functions are obtained by fitting the number of events on each bin in Figures 20, 28 and 41 in~\cite{Abe:2015awa} as well as relevant plots in~\cite{ICHEP-T2K-2016,Salzgeber:2015gua,NOW-T2K-2016}.

\subsection{MINOS}
\label{app:minosformulae}

The $\phi^{r}_{\nu_\alpha}(E_\alpha)$ described in Eq.~(\ref{eq:basic}) is the flux at Far Detector for MINOS. It was obtained by the product of the flux at the Near Detector (ND)  and the Beam Matrix $f^{F/N}({E_\alpha,E_\beta})$, following:
\begin{equation}
 \phi^{r}_{\nu_\alpha}(E_\alpha) = \sum \phi^{(\text{ND})_r}_{\nu_\alpha} (E_{\alpha}) \cdot f^{F/N}({E_\alpha,E_\beta})~,
\end{equation}
where $E_\alpha$ is the bin energy and $E_\beta$ are the adjacent bins of $ E_\alpha$. The ND flux $\phi^{(\text{ND})_r}_{\nu_\alpha} (E_{\alpha})$ was taken from references~\cite{Adamson:2007gu, Adamson:2009ju}. To construct the Beam Matrix $f^{F/N}(E_\alpha, E_\beta)$, we used the matrix $M(E_\alpha, E_\beta)$ extracted from reference \cite{JETP-MINOS-saoulidou-2007}, and also two Gaussian functions
$G_l(G_r)$ with resolutions widths having energy dependence in the maximum until the second order. So, the $f^{F/N}({E_\alpha,E_\beta})$ can be described as:
\begin{equation}
 f^{F/N}({E_\alpha,E_\beta}) = \frac{G_l(E_\alpha, E_\beta) + G_r(E_\alpha,E_\beta)}{2} \times M(E_\alpha, E_\beta)\label{BeamMatrix1}~.
\end{equation}
The functions $G_l(G_r)$ is responsible for extrapolate the flux of higher (lower) to lower (higher) energies.

In MINOS, the only considered interaction is CC. The cross sections $\sigma^{s,\rm CC}_\beta$  are taken from references \cite{Kyberd:2012iz,Hewett:2012ns}. We have taken $E^{\rm CC}_{\mu,\rm rec} = E_{\rm bin}$, and assume the resolution widths $\varsigma^{\rm CC}_\beta(E_\beta)$ to be linear in energy. The efficiency $\epsilon^{CC}_\beta$ of the detector was extracted from reference \cite{Adamson:2007gu}. 
 
\bibliographystyle{JHEP}
\bibliography{nudecay} 

\end{document}